%% file: ungar03_loops07.tex
\documentclass[11pt, a4paper,twoside]{article}
\usepackage{amssymb}
\usepackage{amsmath}
\usepackage{fancyhdr}
\usepackage{amsthm}
\usepackage{rotate}
\usepackage{graphicx}
\usepackage{psfrag}
\usepackage[T1]{fontenc}
\usepackage{afterpage}  %

\newtheorem{definition}{Definition}

\begin{document}
\addtocounter{page}{-10}
%\input{bop-hax.tex}
%\mirror

%\afterpage{\rhead[]{\thepage} \chead[\small A. B. First and C. D.         %%%%%%%%% complete
%Second Author]{\small  Grouplike Loops} \lhead[\thepage]{Gyrogroups} }                  %%%%%%%%% complete
 \afterpage{\rhead[]{\thepage} \chead[\small Abraham A.~Ungar]
 {\small  Grouplike Loops} \lhead[\thepage]{Gyrogroups} }

%\input ../../defs
%%%%%%%%%%%%%%%%%%%%%%%%%%%%%%%%%%%%%%%%%%%%%%%%%%%%%%%%%%%
\newcommand {\Ab}{\mathbf{A}}  % AA insted of Ab produces error msg.
\newcommand {\Bb}{\mathbf{B}}
\newcommand {\Cb}{\mathbf{C}}
\newcommand {\Pb}{\mathbf{P}}
\newcommand {\Rb}{\mathbb{R}}
\newcommand {\Ib}{\mathbb{I}}
\newcommand {\mb}{\mathbf{m}}
\newcommand {\op}{\mathbf{\oplus}} % op has spacing nicer than oplus
\newcommand {\om}{\mathbf{\ominus}} % om has spacing nicer than oplus
\newcommand {\ope}{\op_{_{\subE}}\!\,}
\newcommand {\ode}{\od_{_{\subE}}\!\,}
\newcommand {\ome}{\om_{_{\subE}}\!\,}
\newcommand {\opm}{\op_{_{\subM}}\!\,}
\newcommand {\omm}{\om_{_{\subM}}\!\,}
\newcommand {\od}{\mathbf{\otimes}}   % od has spacing nicer than oplus
\newcommand {\sqp}{\boxplus}
\newcommand {\sqm}{\boxminus}
\newcommand {\gyrab}{\gyr[a,b]}
\newcommand {\gyr}{{\rm gyr}}
\newcommand {\gyrabb}{\gyr[\ab,\bb]}
\newcommand {\Aut}{{\textstyle Aut}}
\newcommand {\ccdot}{\mathbf{\cdot }}
\newcommand {\ab}{\mathbf{a}}
\newcommand {\bb}{\mathbf{b}}
\newcommand {\cb}{\mathbf{c}}
\newcommand {\pb}{\mathbf{p}}
\newcommand {\ub}{\mathbf{u}}
\newcommand {\vb}{\mathbf{v}}
\newcommand {\xb}{\mathbf{x}}
\newcommand {\yb}{\mathbf{y}}
\newcommand {\zb}{\mathbf{z}}
\newcommand {\wb}{\mathbf{w}}
\newcommand {\bz}{\mathbf{0}}
\newcommand {\ro}{r_{_1}}
\newcommand {\rt}{r_{_2}}
\newcommand {\vc}{\mathbb{V}_c}
\newcommand {\vs}{\mathbb{V}_s}
\newcommand {\vsu}{\mathbb{V}_{s=1}}
\newcommand {\vi}{\mathbb{V}}
\newcommand {\BB}{\mathbb{B}}
\newcommand {\inn}{\hspace{-0.1cm}\in\hspace{-0.1cm}}
\newcommand {\gab}{g_{a,b}}
\newcommand {\gub}{\gamma_{\ub}^{\phantom{1}}}
\newcommand {\gvb}{\gamma_{\vb}^{\phantom{1}}}
\newcommand {\gwb}{\gamma_{\wb}^{\phantom{1}}}
\newcommand {\gabs}{\gamma_{\ab}^2}
\newcommand {\gbbs}{\gamma_{\bb}^2}
\newcommand {\gubs}{\gamma_{\ub}^2}
\newcommand {\gvbs}{\gamma_{\vb}^2}
\newcommand {\gwbs}{\gamma_{\wb}^2}
\newcommand {\zerb}{\mathbf{0}}
\newcommand {\half}{\textstyle\frac{1}{2}}
\newcommand {\gabb}{\gamma_{\ab}^{\phantom{1}}}
\newcommand {\gbbb}{\gamma_{\bb}^{\phantom{1}}}
\newcommand {\gcbb}{\gamma_{\cb}^{\phantom{1}}}
\newcommand {\gdbb}{\gamma_{\db}^{\phantom{1}}}
\newcommand {\Rt}{\Rb^3}
\newcommand {\Rct}{{\Rb}_{c}^{3}}
\newcommand {\Rst}{{\Rb}_{s}^{3}}
\newcommand {\Rsn}{{\Rb}_{s}^{n}}
\newcommand {\Rstwo}{{\Rb}_{s}^2}
\newcommand {\uvc}{\displaystyle\frac{\lower.6ex \hbox {$\ub\ccdot\vb$}}{c^2}}
\newcommand {\uvs}{\displaystyle\frac{\lower.6ex \hbox {$\ub\ccdot\vb$}}{s^2}}
\newcommand {\unpuvc}{ \lower.6ex \hbox {$1 + \uvc$} }
\newcommand {\unpuvs}{ \lower.6ex \hbox {$1 + \uvs$} }
\newcommand {\timess}{\!\times\!}
\newcommand {\gupvb}{\gamma_{\ub\op\vb}^{\phantom{1}}}
\newcommand {\opec}{\op_{\subEC}\!\,}
\newcommand {\subEC}{\!\lower.1ex \hbox {\tiny EC}}
\newcommand {\gumvb}{\gamma_{\ub\om\vb}^{\phantom{1}}}
\newcommand {\hb}{\mathbf{h}}
\newcommand {\CC}{\mathbb{C}}
\newcommand {\DD}{\mathbb{D}}
\newcommand {\subE}{\!\lower.1ex \hbox {\tiny E}}
\newcommand {\subM}{\!\lower.1ex \hbox {\tiny M}}
\newcommand {\gyrba}{\gyr[b,a]}
\newcommand {\gyrbab}{\gyr[\bb,\ab]}
\newcommand {\Rtwo}{\Rb^2}
\newcommand {\Rstwou}{{\Rb}_{s=1}^2}
\newcommand {\Db}{\mathbb{D}}
\newcommand {\odm}{\od_{_{\subM}}\!\,}
\newcommand {\subbE}{\!\lower.01ex \hbox {\tiny E}}
\newcommand {\subbM}{\!\lower.01ex \hbox {\tiny M}}
\newcommand {\sqpm}{\sqp_{_{\,\subbM}}\!\,}
\newcommand {\sqmm}{\sqm_{_{\,\subbM}}\!\,}
\newcommand {\NN}{\mathbb{N}}
\newcommand {\gupvbe}{\gamma_{\ub\ope\vb}^{\phantom{1}}}
\newcommand {\sqpe}{\sqp_{_{\,\subbE}}\!\,}
\newcommand {\sqme}{\sqm_{_{\,\subbE}}\!\,}
\newcommand {\Rsthreeu}{{\Rb}_{s=1}^3}
\newcommand {\vbo}{\vb_{\!{_0}}}
%%%%%%%%%%%%%%%%%%%%%%%%%%%%%%%%%%%%%%%%%%%%%%%%%%%%%%%%%%%

\begin{center}
\vspace*{2pt}
{\Large \textbf{Gyrogroups, the Grouplike Loops}}\\[3mm]
{\Large \textbf{in the Service of Hyperbolic Geometry and}}\\[3mm]
{\Large\textbf{Einstein's Special Theory of Relativity}}\\[36pt]
{\large \textsf{\emph{Abraham A.~Ungar}}}
\\[36pt]
\textbf{Abstract}
\end{center}
{\footnotesize
In this era of an increased interest in loop theory,
the Einstein velocity addition law has fresh resonance.
One of the most fascinating aspects of recent work in
Einstein's special theory of relativity is the emergence of
special grouplike loops.
The special grouplike loops, known as {\it gyrocommutative gyrogroups},
have thrust the Einstein velocity addition law,
which previously has operated mostly in the shadows, into the spotlight.
We will find that Einstein (M\"obius) addition is a
gyrocommutative gyrogroup operation that forms the setting for the
Beltrami-Klein (Poincar\'e) ball model of hyperbolic geometry just as
the common vector addition is a commutative group operation that
forms the setting for the standard model of Euclidean geometry.
The resulting analogies to which the grouplike loops give rise lead us
to new results in (i) hyperbolic geometry;
(ii) relativistic physics; and (iii) quantum information and computation.
}



 \enlargethispage{16pt}
% SECTION 1
\section*{\centerline{1. Introduction}}

The author's two recent books with the ambitious titles,
%%%%%%%%%%%%%%%%%%%%%%%%%%%%%%%%%%%%%%%%%%%%%%%%%%%%%%%%%%%%%%%%%%%%%%%%%%%%%
{\it ``Analytic hyperbolic geometry: Mathematical foundations and applications''}
\cite{mybook02}, and
{\it ``Beyond the Einstein addition law and its gyroscopic Thomas
precession: The theory of gyrogroups and gyrovector spaces''}
\cite{mybook01,walterrev2002},
raise expectations for novel applications of special grouplike loops in
hyperbolic geometry and in relativistic physics.
%%%%%%%%%%%%%%%%%%%%%%%%%%%%%%%%%%%%%%%%%%%%%%%%%%%%%%%%%%%%%%%%%%%%%%%%%%%%%
Indeed, these books
lead their readers to see what some special grouplike loops have to offer, and thereby give
them a taste of loops in the service of
the hyperbolic geometry of Bolyai and Lobachevsky and
the special relativity theory of Einstein.

Seemingly structureless, Einstein's relativistic velocity addition
is neither commutative nor associative.
Einstein's failure to recognize and advance the rich,
grouplike loop structure \cite{grouplike}
that regulates his relativistic velocity addition law contributed to
the eclipse of his velocity addition law of relativistically admissible
3-velocities, creating a void that could be filled only with the
Lorentz transformation of 4-velocities, along with its
Minkowski's geometry.

Minkowski characterized his spacetime geometry as evidence that
{\it pre-established harmony} between pure mathematics and applied physics
does exist \cite{pyenson85}.
Subsequently, the study of special relativity
followed the lines laid down by Minkowski, in which the role of
Einstein velocity addition and its interpretation in
the hyperbolic geometry of Bolyai and Lobachevsky
are ignored \cite{barrett98}.
The tension created by the mathematician Minkowski into the specialized realm
of theoretical physics, as well as Minkowski's strategy to overcome
disciplinary obstacles to the acceptance of his reformulation of Einstein's
special relativity is skillfully described by Scott Walter in \cite{walter99a}.

According to Leo Corry \cite{corry98}, Einstein considered Minkowski's
reformulation of his theory in terms of four-dimensional spacetime
to be no more than ``superfluous erudition''.
Admitting that, unlike his seemingly structureless
relativistic velocity addition law,
the Lorentz transformation is an elegant group operation,
Einstein is quoted as saying:

\begin{quotation}
``If you are out to describe truth, leave elegance to the tailor.''
\begin{flushright}
Albert Einstein (1879\,--\,1955)
\end{flushright}
\end{quotation}

One might, therefore, suppose that there is a price to pay
in mathematical elegance and regularity
when replacing
ordinary vector addition approach to Euclidean geometry
with Einstein vector addition approach to hyperbolic geometry.
But,  this is not the case since grouplike loops, called
{\it gyrocommutative gyrogroups}, come to the rescue.
It turns out that Einstein addition of vectors with magnitudes $<c$ is a
gyrocommutative gyrogroup operation and, as such,
it possesses a rich nonassociative algebraic and geometric structure.
The best way to introduce the gyrocommutative gyrogroup notion
that regulates the algebra of Einstein's relativistic velocity addition law
is offered by M\"obius transformations of the disc \cite{kinyonungar00}.
The subsequent transition from M\"obius addition,
which regulates the Poincar\'e ball model of hyperbolic geometry,
Fig.~1, to Einstein addition,
which regulates the Beltrami-Klein ball model of hyperbolic geometry,
Fig.~6, expressed in {gyrolanguage},
will then turn out to be remarkably simple and elegant
\cite{mybook02,ungardiff05}.

Evidently, the grouplike loops that we naturally call
{\it gyrocommutative gyrogroups}, along with their extension to
gyrovector spaces, form a new tool for the twenty-first century
exploration of classical hyperbolic geometry and its use in physics.

%SECTION 2
\section*{\centerline{2. M\"obius transformations of the disc}}

M\"obius transformations of the disc $\DD$,
\begin{equation} \label{osfhe}
\DD= \{z\in \CC :\, |z|<1 \}
\end{equation}
of the complex plane $\CC$
offer an elegant way to introduce
the grouplike loops that we call {\it gyrogroups}.
More than 150 years have passed since August Ferdinand M\"obius
first studied the transformations that now bear his name \cite{wright02}.
Yet, the rich structure he thereby exposed is still far from being
exhausted

Ahlfors' book \cite{ahlfors73},
{\it Conformal Invariants: Topics in Geometric Function Theory},
begins with a presentation of the
M\"obius self-transformation
of the complex open unit disc $\DD$,
\begin{equation} \label{eq01}
z \mapsto e^{i\theta} \frac{a+z}{1+\overline{a}z}=e^{i\theta} (a\opm z)
\end{equation}
$a,z\inn\DD$, $\theta\inn\Rb$,
where $\overline{a}$ is the complex conjugate of $a$
\cite[p.~211]{fisher99}
\cite[p.~185]{greenkrantz}
\cite[pp.~177\,--\,178]{needham97}.
Suggestively, the {\it polar decomposition} \eqref{eq01} of
M\"obius transformation of the disc gives rise to {\it M\"obius addition}, $\opm$,
\begin{equation} \label{eq02}
a\opm z=\frac{a+z}{1+\overline{a}z}
\,.
\end{equation}
Naturally, M\"obius subtraction, $\omm$, is given by $a\omm z = a\opm (-z)$,
so that $z\omm z=0$ and $\omm z=0\omm z=0\opm(-z)=-z$.
Remarkably,
M\"obius addition possesses the
{\it automorphic inverse property}
\begin{equation} \label{eqphic}
\omm (a\opm b) = \omm a\omm b
\end{equation}
and the {\it left cancellation law}
\begin{equation} \label{eq02a}
\omm a\opm(a\opm z)=z
\end{equation}
for all $a,b,z\inn\DD$, \cite{mybook02,mybook01}.

M\"obius addition gives rise to the M\"obius disc groupoid $(\DD,\opm)$,
recalling that a groupoid $(G,\op)$ is a nonempty set, $G$, with
a binary operation, $\op$, and that an automorphism of a
groupoid $(G,\op)$ is a bijective self map $f$ of $G$ that respects its
binary operation $\op$, that is, $f(a\op b)=f(a)\op f(b)$.
The set of all automorphisms of a groupoid $(G,\op)$ forms
a group, denoted $\Aut (G,\op)$.

M\"obius addition $\opm$ in the disc is neither commutative nor associative.
To measure the extent to which M\"obius addition
deviates from associativity we define the {\it gyrator}
\begin{equation} \label{eq07b}
\gyr: \DD\times\DD \rightarrow \Aut(\DD,\opm)
\end{equation}
by the equation
\begin{equation} \label{eq07a}
\gyr[a,b]z = \omm(a\opm b) \opm \{a\opm(b\opm z)\}
\end{equation}
for all $a,b,z\inn\DD$.

The automorphisms
\begin{equation} \label{usjk}
\gyr[a,b] \in Aut(\DD,\opm)
\end{equation}
of $\DD$,
$a,b\inn\DD$, called gyrations of $\DD$, have an important hyperbolic
geometric interpretation \cite{vermeer05}.
Thus, the gyrator in \eqref{eq07b} generates the gyrations in \eqref{usjk}.
In order to emphasize that
gyrations of $\DD$ are also automorphisms of $(\DD,\opm)$,
as we will see below, they are also called {\it gyroautomorphisms}.

Clearly, in the special case when the binary operation $\opm$ in
\eqref{eq07a} is associative, $\gyr[a,b]$ reduces to the trivial
automorphism, $\gyr[a,b]z=z$ for all $z\inn\DD$. Hence, indeed,
the self map $\gyr[a,b]$ of the disc $\DD$ measures the extent
to which M\"obius addition $\opm$ in the disc $\DD$
deviates from associativity.

One can readily simplify \eqref{eq07a} in terms of \eqref{eq02},
obtaining
\begin{equation} \label{eq03h}
\gyrab z= \frac{1+a\overline{b}}{1+\overline{a}b} z
\end{equation}
$a,b,z\inn\DD$, so that the gyrations
\begin{equation} \label{eq03}
\gyrab = \frac{1+a\overline{b}}{1+\overline{a}b}
= \frac{a\opm b}{b\opm a}
\end{equation}
are unimodular complex numbers. As such, gyrations represent
rotations of the disc $\DD$ about its center, as shown in \eqref{eq03h}.

Gyrations are invertible.
The inverse, $\gyr^{-1}[a,b]=(\gyr[a,b])^{-1}$, of a gyration $\gyr[a,b]$
is the gyration $\gyr[b,a]$,
\begin{equation} \label{eq05}
\gyr^{-1}[a,b] = \gyr[b,a]
\end{equation}
Moreover, gyrations respect M\"obius addition in the disc,
\begin{equation} \label{eq05h}
\gyr[a,b](c\opm d)=\gyr[a,b]c \opm \gyr[a,b]d
\end{equation}
for all $a,b,c,d\inn\DD$, so that gyrations of the disc are
automorphisms of the disc, as anticipated in \eqref{usjk}.

Identity \eqref{eq03} can be written as
\begin{equation} \label{eq03a}
a\opm b = \gyrab (b\opm a)
\end{equation}
thus giving rise to the
{\it gyrocommutative law}
of M\"obius addition. Furthermore, Identity \eqref{eq07a} can be
manipulated, by mean of the left cancellation law \eqref{eq02a},
into the identity
\begin{equation} \label{eq03ag}
a\opm(b\opm z) = (a\opm b)\opm \gyrab z
\end{equation}
thus giving rise to the
{\it left gyroassociative law}
of M\"obius addition.

The gyrocommutative law, \eqref{eq03a}, and the
left gyroassociative law, \eqref{eq03ag}, of M\"obius addition in the disc
reveal the grouplike structure of M\"obius groupoid $(\DD,\opm)$, that we
naturally call a {\it gyrocommutative gyrogroup.}
Taking the key features of M\"obius groupoid $(\DD,\opm)$ as axioms,
and guided by analogies with group theory, we
thus obtain the following definitions of gyrogroups and
gyrocommutative gyrogroups.

% DEFINITION NUMBER 1
\begin{definition}\label{defroupx}
{\bf (Gyrogroups).}
{\it
A groupoid $(G , \op )$
is a gyrogroup if its binary operation satisfies the following axioms.
In $G$ there is at least one element, $0$, called a left identity, satisfying

\noindent
(G1) \hspace{1.2cm} $0 \op a=a$

\noindent
for all $a \in G$. There is an element $0 \in G$ satisfying axiom $(G1)$ such
that for each $a\in G$ there is an element $\om a\in G$, called a
left inverse of $a$, satisfying

\noindent
(G2) \hspace{1.2cm} $\om a \op a=0\,.$

\noindent
Moreover, for any $a,b,c\in G$ there exists a unique element $\gyr[a,b]c \in G$
such that the binary operation obeys the left gyroassociative law

\noindent
(G3) \hspace{1.2cm} $a\op(b\op c)=(a\op b)\op\gyrab c\,.$

\noindent
The map $\gyr[a,b]:G\to G$ given by $c\mapsto \gyr[a,b]c$
is an automorphism of the groupoid $(G,\op)$, that is,

\noindent
(G4) \hspace{1.2cm} $\gyrab\in\Aut (G,\op) \,,$

\noindent
and the automorphism $\gyr[a,b]$ of $G$ is called
the gyroautomorphism, or the gyration, of $G$ generated by $a,b \in G$.
The operator $\gyr : G\times G\rightarrow\Aut (G,\op)$ is called the
gyrator of $G$.
Finally, the gyroautomorphism $\gyr[a,b]$ generated by any $a,b \in G$
possesses the left loop property

\noindent
(G5) \hspace{1.2cm} $\gyrab=\gyr [a\op b,b] \,.$
}
\end{definition}

The gyrogroup axioms ($G1$)\,--\,($G5$)
in Definition \ref{defroupx} are classified into three classes:
%%%%%%%%%%%%%%%%%%%%%%%%%%%%%%%%%%%%%%%%%%%%%%%%%%%%%%%%%%%%%%%%%%%%%%%%%%%%%
\begin{itemize}
\item[$(1)$]
The first pair of axioms, $(G1)$ and $(G2)$, is a reminiscent of the
group axioms.
\item[$(2)$]
The last pair of axioms, $(G4)$ and $(G5)$, presents the gyrator
axioms.
\item[$(3)$]
The middle axiom, $(G3)$, is a hybrid axiom linking the two pairs of
axioms in (1) and (2).
\end{itemize}
%%%%%%%%%%%%%%%%%%%%%%%%%%%%%%%%%%%%%%%%%%%%%%%%%%%%%%%%%%%%%%%%%%%%%%%%%%%%%

The loop property $(G5)$ turns out to be equivalent to the
gyration-free identity
\begin{equation} \label{gjdn}
x\op(y\op(x\op z)) = (x\op(y\op x))\op z
\end{equation}
which loop theorists recognize as the {\it left Bol identity}
\cite{sabinin95,sabinin98}.

As in group theory, we use the notation
$a \om b = a \op (\om b)$
in gyrogroup theory as well.

In full analogy with groups, gyrogroups are classified into gyrocommutative and
non-gyrocommutative gyrogroups.

% DEFINITION NUMBER 2
\begin{definition}\label{defgyrocomm}
{\bf (Gyrocommutative Gyrogroups).}
{\it
A gyrogroup $(G, \oplus )$ is gyrocommutative if
its binary operation obeys the gyrocommutative law

\noindent
(G6) \hspace{1.2cm} $a\oplus b=\gyrab(b\oplus a)$

\noindent
for all $a,b\in G$.
}
\end{definition}
%%%%%%%%%%%%%%%%%%%%%%%%%%%%%%%%%%%%%%%%%%%%%%%%%%%%%%%

Some first gyrogroup theorems, some of which are analogous to
group theorems, are presented in \cite[Chap.~2]{mybook02}.
Thus, in particular, the gyrogroup left identity and left inverse
are identical with their right counterparts, and the resulting
identity and inverse are unique, as in group theory.
Furthermore, the left gyroassociative law and the left loop property
are associated with corresponding right counterparts.

A gyrogroup operation $\op$ comes with a dual operation,
the {\it cooperation} (or, {\it co-operation}, for clarity)
$\sqp$ \cite[Def.~2.7]{mybook02},
given by the equation
\begin{equation} \label{eqdfhn01}
a\sqp b = a \op\gyr[a,\om b]b
\end{equation}
so that
\begin{equation} \label{eqdfhn02}
a\sqm b = a \om\gyr[a,b]b
\end{equation}
for all $a,b\in G$,
where we define $a\sqm b = a\sqp (\om b)$.
The gyrogroup cooperation shares with its associated gyrogroup operation
remarkable duality symmetries as, for instance
\cite[Theorem 2.10]{mybook02},
%%%%%%%%%%%%%%%%%%%%%%%%%%%%%%%%%%%%%%%%%%%%%%%%%%%%%%%%%%%%%%%%%%%%
\begin{equation} \label{jsnax}
\begin{split}
a \sqp b &= a \op  \gyr [a,\om b]b \\
a \op  b &= a \sqp \gyr [a,b]b
\end{split}
\end{equation}
%%%%%%%%%%%%%%%%%%%%%%%%%%%%%%%%%%%%%%%%%%%%%%%%%%%%%%%%%%%%%%%%%%%%

Interestingly, by \cite[Theorem 3.4]{mybook02},
a gyrogroup cooperation is commutative if and only if its
corresponding gyrogroup is gyrocommutative.
%%%%%%%%%%%%%%%%%%%%%%%%%%%%%%%%%%%%%%%%%%%%%%%%%%%%%%%

The gyroautomorphisms have their own rich structure
as we see, for instance, from the gyroautomorphism inversion property
\begin{equation} \label{eq0disc04a}
(\gyrab)^{-1} = \gyrba
\end{equation}
from the loop property (left and right)
\begin{equation} \label{eq0disc05}
 \begin{split}
\gyrab &= \gyr[a\op b, b] \\
\gyrab &= \gyr[a,b\op a]
 \end{split}
 \end{equation}
and from the elegant nested gyroautomorphism identity
\begin{equation} \label{eq0disc06}
\gyrab = \gyr[\om\gyrab b,a]
\end{equation}
for all $a,b\in G$ in any gyrogroup $G=(G,\op)$.
More gyroautomorphism identities and important
gyrogroup theorems, along with their applications,
are found in \cite{mybook01,mybook02,mybook03} and in
\cite{gracielaungar01,feder03,nourou99,nouro01,kuznetsov03,rozga00,
sabinin95,sabinin98,vermeer05}.
%%%%%%%%%%%%%%%%%%%%%%%%%%%%%%%%%%%%%%%%%%%%%%%%%%%%%%%

Thus, without losing the flavor of the group structure
we have generalized it into the gyrogroup structure to suit
the needs of M\"obius addition in the disc and, more generally,
in the open ball of any real inner product space \cite{mbtogyp07},
as we will show in Sec.~3.
Gyrogroups abound in group theory, as shown
in \cite{tuvalungar01} and \cite{tuvalungar02}, where finite
and infinite gyrogroups, both gyrocommutative and
non-gyrocommutative, are studied. Plenty of gyrogroup theorems
are found in \cite{mybook01,mybook02,mybook03}.
Furthermore, any gyrogroup can be extended into a group,
called a {\it gyrosemidirect product group}
\cite[Sec.~2.6]{mybook02} \cite{kinyonjones00}.
Hence, the generalization of groups into gyrogroups
bears an intriguing resemblance to the generalization of the rational numbers
into the real ones. The beginner is initially surprised to discover an
irrational number, like $\sqrt{2}$, but soon later
he is likely to realize
that there are more irrational numbers than rational ones.
Similarly, the gyrogroup structure of M\"obius addition initially comes
as a surprise.
But, interested explorers may soon realize that in some sense
there are more non-group gyrogroups than groups.

In our ``gyrolanguage'', as the reader has noticed, we attach the prefix
``gyro'' to a classical term to mean the analogous term in
our study of grouplike loops.
The prefix stems from Thomas gyration, which is the mathematical
abstraction of the relativistic effect known as Thomas precession,
explained in \cite{mybook01}.
Indeed, gyrolanguage turns out to be the language we need
to articulate novel analogies that the classical and the modern
in this paper and in \cite{mybook01,mybook02,mybook03} share.

%SECTION 3
\section*{\centerline{3. M\"obius Addition in the Ball}}

If we identify complex numbers of the complex plane $\CC$ with
vectors of the Euclidean plane $\Rtwo$ in the usual way,
\begin{equation} \label{eq6100}
\CC \ni u = u_1+iu_2 = (u_1,u_2) = \ub \in \Rtwo
\end{equation}
then the inner product and the norm in $\Rtwo$ are given by
the equations
\begin{equation} \label{eq6101}
\begin{array}{c}
\bar{u}v+u\bar{v} = 2\ub\ccdot\vb
\\[8pt]
|u| = \|\ub\|
\end{array}
\end{equation}
These, in turn, enable us to translate M\"obius addition from the complex open
unit disc $\DD$ into the open unit disc
$\Rstwou=\{\vb\inn\Rtwo:\|\vb\|<s=1\}$ of $\Rtwo$ \cite{kinyonungar00}:
%%%%%%%%%%%%%%%%%%%%%%%%%%%%%%%%%%%%%%%%%%%%%%%%%%%%%%%%%%%%%%%%%%%%
\begin{equation} \label{eq6102}
\begin{split}
\DD \ni
u\opm v &= \frac{u+v}{1+\bar{u}v}                      \\
&= \frac{(1+u\bar{v})(u+v)} {(1+\bar{u}v)(1+u\bar{v})}\\
&= \frac{(1+\bar{u}v+u\bar{v}+|v|^2)u+(1-|u|^2)v}
        {1+\bar{u}v+u\bar{v}+|u|^2|v|^2}              \\
&= \frac{(1+2\ub\ccdot\vb+\|\vb\|^2)\ub+(1-\|\ub\|^2)\vb}
         {1+2\ub\ccdot\vb+\|\ub\|^2\|\vb\|^2}         \\
&= \ub\opm\vb \in \Rstwou
\end{split}
\end{equation}
%%%%%%%%%%%%%%%%%%%%%%%%%%%%%%%%%%%%%%%%%%%%%%%%%%%%%%%%%%%%%%%%%%%%
for all $u,v\in\Db$ and all $\ub,\vb\in\Rstwou$.
The last equation in \eqref{eq6102} is a vector equation, so that
its restriction to the ball of the Euclidean two-dimensional space
$\Rstwou$ is a mere artifact.
As such, it survives unimpaired in higher dimensions,
suggesting the following
definition of M\"obius addition in the ball of any real inner
product space.

%%%%%%%%%%%%%%%%%%%%%%%%%%%%%%%%%%%%%%%%%%%%%%%%%%%%%%%%%%%%%%%%%%%%
% DEFINITION NUMBER 3
\begin{definition}\label{defmobiusadd}
{\bf (M\"obius Addition in the Ball).}
{\it
Let $\vi$ be a real inner product space \cite{marsden74}, and let
$\vs$ be the $s$-ball of $\vi$,
\begin{equation} \label{eqsball}
\vs = \{\vs\in\vi : \|\vb\|<s\}
\end{equation}
for any fixed $s>0$.
M\"obius addition $\opm$ in the ball $\vs$
is a binary operation in $\vs$ given by
the equation
%%%%%%%%%%%%%%%%%%%%%%%%%%%%%%%%%%%%%%%%%%%%%%%%%%%%%%%%%%%%%%%%%%%%
 \begin{equation} \label{eq696}
\ub\opm\vb
 = \frac{(1+\frac{2}{s^2}\ub\ccdot\vb+\frac{1}{s^2}\|\vb\|^2 )\ub
        +(1-\frac{1}{s^2}\|\ub\|^2)\vb}
        {1+\frac{2}{s^2}\ub\ccdot\vb+\frac{1}{s^4}\|\ub\|^2\|\vb\|^2}
 \end{equation}
$\ub,\vb\inn\vs$,
where $\ccdot$ and $\|\ccdot\|$ are the inner product and norm that the
ball $\vs$ inherits from its space $\vi$.
}
\end{definition}
%%%%%%%%%%%%%%%%%%%%%%%%%%%%%%%%%%%%%%%%%%%%%%%%%%%%%%%%%%%%%%%%%%%%

Without loss of generality, one may select $s=1$ in
Definition \ref{defmobiusadd}.
We, however, prefer to keep $s$ as a free positive parameter in order
to exhibit the result that in the limit as $s\rightarrow\infty$,
the ball $\vs$ expands to the whole of its real inner product space $\vi$,
and M\"obius addition $\opm$ in the ball reduces to vector addition in
the space.
Remarkably, like the M\"obius disc groupoid $(\DD,\opm)$,
also the M\"obius ball groupoid
$(\vs,\opm)$ forms a gyrocommutative gyrogroup,
called a M\"obius gyrogroup.

M\"obius addition in the ball $\vs$ is known in the literature
as a {\it hyperbolic translation} \cite{ahlfors81,ratcliffe94}.
Following the discovery of the gyrocommutative gyrogroup structure
in 1988 \cite{parametrization}, M\"obius hyperbolic translation in
the ball $\vs$ now deserves the title
{\it ``M\"obius addition''} in the ball $\vs$, in full analogy with the
standard vector addition in the space $\vi$ that contains the ball.

M\"obius addition in the ball $\vs$ satisfies the {\it gamma identity}
\begin{equation} \label{eq6gupv72}
\gamma_{\ub\opm\vb}^{\phantom{O}}
= \gub\gvb\sqrt
{1+\frac{2}{s^2}\ub\ccdot\vb+\frac{1}{s^4}\|\ub\|^2\|\vb\|^2}
\end{equation}
for all $\ub,\vb\in\vs$,
where $\gub$ is the gamma factor
\begin{equation} \label{eqeq6gupv72gs}
\gvb = \frac{1}{\sqrt{1-\displaystyle\frac{\|\vb\|^2}{s^2}}}
\end{equation}
in the $s$-ball $\vs$.

Following \eqref{eqdfhn01}, M\"obius cooperation, also called
M\"obius coaddition, in the ball is
commutative, given by the equation
\begin{equation} \label{eq7mbdual84}
\ub\sqpm\vb = \frac{\gubs\ub+\gvbs\vb} {\gubs+\gvbs-1}
\end{equation}
for all $\ub,\vb\inn\vs$.
Note that $\vb\sqpm\zerb=\vb$ and $\vb\sqmm\vb=\zerb$, as expected.

%SECTION 4
\section*{\centerline{4. Gyrogroups Are Loops}}

A {\it loop} is a groupoid $(G,\op)$ with an identity element, 0,
such that each of its two {\it loop equations} for the unknowns $x$ and $y$,
%%%%%%%%%%%%%%%%%%%%%%%%%%%%%%%%%%%%%%%%%%%%%%%%%%%%%%%%%%%%%%%%%%%%
\begin{equation} \label{uenbdf}
\begin{split}
a\op x &= b\\
y\op a &= b
\end{split}
\end{equation}
%%%%%%%%%%%%%%%%%%%%%%%%%%%%%%%%%%%%%%%%%%%%%%%%%%%%%%%%%%%%%%%%%%%%
possesses a unique solution in $G$ for any $a,b\in G$
\cite{pflugfelder90,pflugfelder00}.
Any gyrogroup is a loop.
Indeed, if $(G,\op)$ is a gyrogroup then
the respective unique solutions of the gyrogroup {\it loop equations}
\eqref{uenbdf} are \cite[Sec.~2.4]{mybook02}
%%%%%%%%%%%%%%%%%%%%%%%%%%%%%%%%%%%%%%%%%%%%%%%%%%%%%%%%%%%%%%%%%%%%
\begin{equation} \label{uendfh1}
\begin{split}
x &= \om a\op b \\
y &= b\sqm a
\end{split}
\end{equation}
%%%%%%%%%%%%%%%%%%%%%%%%%%%%%%%%%%%%%%%%%%%%%%%%%%%%%%%%%%%%%%%%%%%%

The {\it cogyrogroup} $(G,\sqp)$, associated with any gyrogroup $(G,\op)$,
is also a loop. The unique solutions of its two loop equations
%%%%%%%%%%%%%%%%%%%%%%%%%%%%%%%%%%%%%%%%%%%%%%%%%%%%%%%%%%%%%%%%%%%%
\begin{equation} \label{uenkdf}
\begin{split}
a\sqp x &= b\\
y\sqp a &= b
\end{split}
\end{equation}
%%%%%%%%%%%%%%%%%%%%%%%%%%%%%%%%%%%%%%%%%%%%%%%%%%%%%%%%%%%%%%%%%%%%
are \cite[Theorem 2.38]{mybook02}
%%%%%%%%%%%%%%%%%%%%%%%%%%%%%%%%%%%%%%%%%%%%%%%%%%%%%%%%%%%%%%%%%%%%
\begin{equation} \label{ujndfh1}
\begin{split}
x &= \om(\om b\op a) \\
y &= b\om a
\end{split}
\end{equation}
%%%%%%%%%%%%%%%%%%%%%%%%%%%%%%%%%%%%%%%%%%%%%%%%%%%%%%%%%%%%%%%%%%%%

Note that, in general, the two loop equations in \eqref{uenkdf} are
identically the same equation if and only if the gyrogroup cooperation
$\sqp$ is commutative. Hence, their solutions must be, in general,
identical if and only if the gyrogroup cooperation $\sqp$ is commutative.
Indeed,
a gyrogroup $(G,\op)$ possesses the {\it gyroautomorphic inverse property},
$\om(a\op b)=\om a\om b$, if and only if it is gyrocommutative
\cite[Theorem 3.2]{mybook02}.
Hence, the two solutions, $x$ and $y$, in \eqref{ujndfh1} are, in general,
equal if and only if the gyrogroup $(G,\op)$ is gyrocommutative. This result
is compatible with the result that a gyrogroup is gyrocommutative
if and only if its cooperation $\sqp$ is commutative
\cite[Theorem 3.4]{mybook02}.

The cogyrogroup is an important and interesting loop.
Its algebraic structure is not grouplike, but it plays a crucial
role in the study of the gyroparallelogram law of Einstein's
special relativity theory and its underlying hyperbolic geometry,
Figs.~4, 5 and 8.

It follows from the solutions of the loop equations in
\eqref{uenbdf} and \eqref{uenkdf} that
any gyrogroup $(G,\op)$ possesses the following
cancellation laws \cite[Table 2.1]{mybook02}:
%%%%%%%%%%%%%%%%%%%%%%%%%%%%%%%%%%%%%%%%%%%%%%%%%%%%%%%%%%%%%%%%%%%%
\begin{equation} \label{uendfh2}
\begin{split}
a\op(\om a\op b) &= b \\
(b\sqm a)\op a &= b \\
a\sqm(\om b\op a) &= b \\
(b\om a)\sqp a &= b \\
\end{split}
\end{equation}
%%%%%%%%%%%%%%%%%%%%%%%%%%%%%%%%%%%%%%%%%%%%%%%%%%%%%%%%%%%%%%%%%%%%
The first (second) cancellation law in \eqref{uendfh2}
is called the {\it left (right) cancellation law}.
The last cancellation law in \eqref{uendfh2} is called
the {\it second right cancellation law}.
The two right cancellation laws in \eqref{uendfh2} form one of the
duality symmetries that the gyrogroup operation and cooperation share,
mentioned in the paragraph of \eqref{jsnax}.
It is thus clear that in order to maintain analogies between
gyrogroups and groups, we need both the gyrogroup operation and its
associated gyrogroup cooperation.

In the special case when a gyrogroup is gyrocommutative, it is also
known as
(i) a {\it K-loop}
(a term coined by Ungar in \cite{noncomm}; see also
\cite[pp.~1,~169-170]{kiechle02}); and
(ii) a {\it Bruck loop} \cite[pp.~168]{kiechle02}.
A new term,
(iii) ``dyadic symset'', which emerges from an interesting work of
Lawson and Lim in \cite{lawson04},
turns out, according to \cite[Theorem 8.8]{lawson04},
to be identical with a
two-divisible, torsion-free, gyrocommutative gyrogroup \cite[p.~71]{mybook02}.

%SECTION 5
\section*{\centerline{5. M\"obius scalar multiplication in the Ball}}

Having developed the M\"obius gyrogroup as a grouplike loop,
we do not stop at the
loop level. Encouraged by analogies gyrogroups share with groups,
we now seek analogies with vector spaces as well.
Accordingly, we uncover the scalar multiplication, $\odm$,
between a real number $r\inn\Rb$ and a vector $\vb\inn\vs$,
that a M\"obius gyrogroup $(\vs,\opm)$ admits, so that we can
turn the M\"obius gyrogroup into a M\"obius gyrovector space $(\vs,\opm,\odm)$.
For any natural number $n\inn\NN$ we define and calculate
$n\odm\vb := \vb \opm \, \ldots \,\opm\vb$ ($n$-terms), obtaining a result
in which we formally replace $n$ by a real number $r$, suggesting
the following definition of
the M\"obius scalar multiplication.
%%%%%%%%%%%%%%%%%%%%%%%%%%%%%%%%%%%%%%%%%%%%%%%%%%%%%%%%%%%%%%%%%%%%
% DEFINITION NUMBER 6.80
\begin{definition}\label{defmbscalar48} 
{\bf (M\"obius Scalar Multiplication).}
{\it
Let $(\vs,\opm)$ be a M\"obius gyrogroup.
Then its corresponding M\"obius gyrovector space $(\vs,\opm,\odm)$ involves
the M\"obius scalar multiplication $r\odm\vb=\vb\odm r$ in $\vs$,
given by the equation
 \begin{equation} \label{eqmlt03}
 \begin{split}
r\odm\vb&= s \frac
 {\left(1+\displaystyle\frac{\|\vb\|}{s}\right)^r 
- \left(1-\displaystyle\frac{\|\vb\|}{s}\right)^r}
 {\left(1+\displaystyle\frac{\|\vb\|}{s}\right)^r 
+ \left(1-\displaystyle\frac{\|\vb\|}{s}\right)^r}
\frac{\vb}{\|\vb\|}\\[8pt]
&= s \tanh( r\,\tanh^{-1}\frac{\|\vb\|}{s})\frac{\vb}{\|\vb\|}
 \end{split}
 \end{equation}
where $r\inn\Rb$, $\vb\inn\vs$, $\vb\ne\bz$; and $r\odm\bz=\bz$.
}
\end{definition}
%%%%%%%%%%%%%%%%%%%%%%%%%%%%%%%%%%%%%%%%%%%%%%%%%%%%%%%%%%%%%%%%%%%%

Extending Def.~\ref{defmbscalar48} by abstraction, we obtain the
abstract gyrovector space, studied in \cite[Chap.~6]{mybook02}.
As we go through the study of gyrovector spaces, we see remarkable
analogies with classical results unfolding. In particular,
armed with the gyrovector space structure, we offer a
gyrovector space approach to the study of hyperbolic geometry
\cite{mybook02}, which is fully analogous to the common
vector space approach to the study of Euclidean geometry
\cite{hausner98}. Our basic examples are presented in the sequel
and shown in several figures.

%SECTION 6
\section*{\centerline{6. M\"obius Gyroline and More}}

In full analogy with straight lines in the standard vector space approach to
Euclidean geometry, let us consider the
{\it gyroline} equation in the ball $\vs$,
\begin{equation} \label{rjsk}
L_{^{AB}} := A \op (\om A \op B)\od t
\end{equation}
$t\inn\Rb$, $A,B\in\vs$,
in a M\"obius gyrovector space $(\vs,\op,\od)$.
For simplicity, we use in this section the notation
$\opm=\op$ and $\odm=\od$.
The gyrosegment $AB$ is the part of the gyroline \eqref{rjsk} that links
the points $A$ and $B$. Hence, it is given by \eqref{rjsk} with $0\le t\le1$,
Fig.~1.

For any $t\inn\Rb$ the point $P(t)=A \op (\om A \op B)\od t$
lies on the gyroline $L_{^{AB}}$.
Thinking of $t$ as time, at time $t=0$ the point $P$ lies at $P(0)=A$
and, owing to the left cancellation law in \eqref{uendfh2},
at time $t=1$ the point $P$ lies at $P(1)=B$.
Furthermore, the point $P$ reaches the {\it gyromidpoint} $M_{^{AB}}$
of the points $A$ and $B$ at time $t=1/2$,
\begin{equation} \label{ysmha}
M_{^{AB}} = A \op (\om A \op B)\od \half = \half\od(A\sqp B)
\end{equation}
\cite[Sec.~6.5]{mybook02}.
Here $M_{^{AB}}$ is the unique gyromidpoint of the points $A$ and $B$
in the gyrodistance sense, $d(A,M_{^{AB}}) = d(B,M_{^{AB}})$,
the gyrodistance function being $d(A,B)=\|\om A\op B\|=\|B\om A\|$.

%%%%%%%%%%%%%%%%%%%%%%%%%%%%%%%%%%%%%%%%%%%%%%%%%%%%%%%%%%%%%%%%%%%%
%FIGURES 1-2: Mobius  gyroline and cogyroline.   Updated: .eps -> .ps
% Updated to Figure 1. Only a Mobius gyroline.
%%%%%%%%%%%%%%%%%%%%%%%%%%%%%%%%%%%%%%%%%%%%%%%%%%%%%%%%%%%%%%%%%%%%
 \input{fig174eps} % The gyrotriangle in (+)M 
%%%%%%%%%%%%%%%%%%%%%%%%%%%%%%%%%%%%%%%%%%%%%%%%%%%%%%%%%%%%%%%%%%%%
%\input{fig163164c.eps} % The gyroline and cogyroline
%       in (+)M  Fig. 1 Fig.~\ref{fig163164cm}
%% \begin{figure}
%% \centerline{\includegraphics[width=10cm]{new_fig163164c.eps}}
%\caption{
%The gyroline (left) and the cogyroline (right) in a
%M\"obius gyrovector space of a disc
%are Euclidean circular arcs that intersect the boundary of the disc,
%respectively, orthogonally and diametrically.
%The gyroline is recognized as the
%geodesic line in the Poincar\'e ball model of hyperbolic geometry
%\cite{ungardiff05}.
%Shown are a generic point $\pb$, and the midpoint $\mb_{\ab\bb}$.
%\label{fig163164cm}}
%% \end{figure}
%%%%%%%%%%%%%%%%%%%%%%%%%%%%%%%%%%%%%%%%%%%%%%%%%%%%%%%%%%%%%%%%%%%%

%%%%%%%%%%%%%%%%%%%%%%%%%%%%%%%%%%%%%%%%%%%%%%%%%%%%%%%%%%%%%%%%%%%%
%FIGURE 3: The Mobius gyrotriangle         Updated: .eps -> .ps
% Updated to fig. 2.
%%%%%%%%%%%%%%%%%%%%%%%%%%%%%%%%%%%%%%%%%%%%%%%%%%%%%%%%%%%%%%%%%%%%
 \input{fig148kps} % The gyrotriangle in (+)M  Fig. 2 Fig.~\ref{fig148km}
%%%%%%%%%%%%%%%%%%%%%%%%%%%%%%%%%%%%%%%%%%%%%%%%%%%%%%%

In the special case when $\vs=\Rstwo$, the gyroline $L_{^{AB}}$, shown in
Fig.~1,
is a circular arc that intersects the boundary
of the $s$-disc $\Rstwo$ orthogonally.
A study of the connection between gyrovector spaces and differential geometry
\cite[Chap.~7]{mybook02} \cite{ungardiff05} reveals that this gyroline
is the unique geodesic that passes through the
points $A$ and $B$ in the Poincar\'e disc model of hyperbolic geometry.

%%%%%%%%%%%%%%%%%%%%%%%%%%%%%%%%%%%%%%%%%%%%%%%%%%%%%%%%%%
The
{\it cogyroline} equation in the ball $\vs$, similar to \eqref{rjsk}, is
\begin{equation} \label{corjsk}
L_{^{AB}}^c := (B \sqm A)\od t \op A
\end{equation}
$t\inn\Rb$, $A,B\in\vs$,
in a M\"obius gyrovector space $(\vs,\op,\od)$.
The cogyrosegment $AB$ is the part of the cogyroline \eqref{corjsk} that links
the points $A$ and $B$. Hence, it is given by \eqref{corjsk} with $0\le t\le1$,
Fig.~2.

For any $t\inn\Rb$ the point $P(t)= (B \sqm A)\od t \op A$
lies on the cogyroline $L_{^{AB}}^c$ in \eqref{corjsk}.
Thinking of $t$ as time, at time $t=0$ the point $P$ lies at $P(0)=A$
and, owing to the right cancellation law in \eqref{uendfh2},
at time $t=1$ the point $P$ lies at $P(1)=B$.
Furthermore, the point $P$ reaches the {\it cogyromidpoint} $M_{^{AB}}^c$
of the points $A$ and $B$ at time $t=1/2$,
\begin{equation} \label{coysmha}
M_{^{AB}}^c = (B \sqm A)\od \half \op A = \half \od (A\op B)
\end{equation}
\cite[Theorem 6.34]{mybook02}.
Here $M_{^{AB}}^c$ is the unique cogyromidpoint of the points $A$ and $B$
in the cogyrodistance sense, $d^c(A,M_{^{AB}}^c) = d^c(B,M_{^{AB}}^c)$,
the cogyrodistance function being $d^c(A,B)=\|\om A\sqp B\|=\|B\sqm A\|$.

%%%%%%%%%%%%%%%%%%%%%%%%%%%%%%%%%%%%%%%%%%%%%%%%%%%%%%%
%FIGURES 4-5: Mobius gyroparallelogram - sidebyside      Deleted.
%%%%%%%%%%%%%%%%%%%%%%%%%%%%%%%%%%%%%%%%%%%%%%%%%%%%%%%
% input figs. 4-5      Fig.~\ref{fig140da1140da2m}
%%%%%%%%%%%%%%%%%%%%%%%%%%%%%%%%%%%%%%%%%%%%%%%%%%%%%%%
%%% \begin{figure}
%%%  \centerline{\includegraphics[width=10cm]{new_fig140da1140da2.eps}}
%\caption{The M\"obius gyroparallelogram
%$ABDC$, $D=(B\sqp C)\om A$,
%and the gyroparallelogram law.
%Any three non-gyrocollinear points, $A,B,C$, in a gyrovector space
%of dimension $\ge2$ generate a gyroparallelogram $ABDC$
%if and only if $D$ satisfies the
%gyroparallelogram condition $D=(B\sqp C)\om A$.
%\label{fig140da1140da2m}}
%%% \end{figure}
%%%%%%%%%%%%%%%%%%%%%%%%%%%%%%%%%%%%%%%%%%%%%%%%%%%%%%%

In the special case when $\vs=\Rstwo$, the
cogyroline $L_{^{AB}}^c$, shown in
Fig.~2, is a circular arc that intersects the boundary
of the $s$-disc $\Rstwo$ diametrically.
%%%%%%%%%%%%%%%%%%%%%%%%%%%%%%%%%%%%%%%%%%%%%%%%%%%%%%%%%%

Let $A,B,C\inn G$ be any three non-gyrocollinear points of a
M\"obius gyrovector space
$G=(G,\op,\od)$.
In Fig.~3 we see a gyrotriangle $ABC$
whose vertices, $A$, $B$, and $C$, are linked by the
gyrovectors $\ab$, $\bb$, and $\cb$; and
whose side gyrolengths are $a$, $b$, and $c$,
given by the equations
 \begin{equation} \label{eq6cd02b}
 \begin{split}
\ab &= \om C  \op  B , \hspace{1.14cm} a = \|\ab\| \\
\bb &= \om C  \op  A , \hspace{1.14cm} b = \|\bb\| \\
\cb &= \om B  \op  A , \hspace{1.16cm} c = \|\cb\|
 \end{split}
 \end{equation}

With the gyrodistance function $d(A,B)=\|\om A\op B\|=\|B\om A\|$,
we have the gyrotriangle inequality \cite[Theorem 6.9]{mybook02}
$d(A,C) \le d(A,B) \op d(B,C)$,
in full analogy with the Euclidean triangle inequality.

A gyrovector $\vb=\om A\op B$ in a M\"obius gyrovector plane
$(\Rstwo,\op,\od)$ and in a M\"obius three-dimensional
gyrovector space $(\Rst,\op,\od)$
is represented graphically by the directed gyrosegment $AB$
from $A$ to $B$ as, for instance, in
Figs.~4\,--\,5 and 8.

Two gyrovectors,
(i) $\om A\op B$, from $A$ to $B$, and
(ii) $\om A^{\prime}\op B^{\prime}$, from $A^{\prime}$ to $B^{\prime}$,
in a gyrovector space $G=(G,\op,\od)$ are equivalent if
\begin{equation} \label{ofjsnb}
\om A\op B = \om A^{\prime}\op B^{\prime}
\end{equation}

In the same way that vectors in Euclidean geometry are
equivalence classes of directed segments
that add according to the parallelogram law,
gyrovectors in hyperbolic geometry are
equivalence classes of directed gyrosegments that add according to the
gyroparallelogram law.
A gyroparallelogram, the hyperbolic parallelogram,
sounds like a contradiction in terms since parallelism in hyperbolic geometry
is denied. However, in full analogy with Euclidean geometry, but with no
reference to parallelism, the gyroparallelogram is defined as a hyperbolic
quadrilateral whose gyrodiagonals intersect at their gyromidpoints,
as in Figs.~4\,--\,5.
Indeed, any three non-gyrocollinear points $A,B,C$ in a gyrovector space
$(G,\op,\od)$ form a gyroparallelogram $ABDC$ if and only if $D$
satisfied the {\it gyroparallelogram condition} $D=(B\sqp C)\om A$
\cite[Sec.~6.7]{mybook02}.

An interesting contrast between Euclidean and hyperbolic geometry
is observed here. In Euclidean geometry vector addition coincides with
the parallelogram addition law.
In contrast, in hyperbolic geometry gyrovector addition, given by
M\"obius addition, and the M\"obius gyroparallelogram addition law are
distinct.

%SECTION 7
\section*{\centerline{7. Einstein Operations in the Ball}}

%%%%%%%%%%%%%%%%%%%%%%%%%%%%%%%%%%%%%%%%%%%%%%%%%%%%%%%%%%%%%%%%%%%%
% DEFINITION NUMBER 5
\begin{definition}\label{defeinstadd}
{\bf (Einstein Addition in the Ball).}
{\it
Let $\vi$ be a real inner product space and let
$\vs$ be the $s$-ball of $\vi$,
\begin{equation} \label{eqsballa}
\vs = \{\vb\in\vi : \|\vb\|<s\}
\end{equation}
where $s\!>\!0$ is an arbitrarily fixed constant (that represents
in physics the vacuum speed of light $c$).
Einstein addition $\ope$ is a binary operation in $\vs$ given by
the equation
%%%%%%%%%%%%%%%%%%%%%%%%%%%%%%%%%%%%%%%%%%%%%%%%%%%%%%%%%%%%%%%%%%%%
\begin{equation} \label{eqEA}                      %Einstein's Addition
{\ub}\ope{\vb}=\frac{1}{\unpuvs}
\left\{ {\ub}+ \frac{1}{\gub}\vb+\frac{1}{s^{2}}\frac{\gamma _{{\ub}}}{%
1+\gamma _{{\ub}}}( {\ub}\ccdot{\vb}) {\ub} \right\}
\end{equation}
where $\gub$ is the gamma factor, \eqref{eqeq6gupv72gs}, in $\vs$, and
where $\ccdot$ and $\|\ccdot\|$ are the inner product and norm that the
ball $\vs$ inherits from its space $\vi$.
}
\end{definition}
%%%%%%%%%%%%%%%%%%%%%%%%%%%%%%%%%%%%%%%%%%%%%%%%%%%%%%%%%%%%%%%%%%%%

We may note that the Euclidean 3-vector algebra was not so
widely known in 1905 and, consequently, was not used by Einstein.
Einstein calculated in his founding paper \cite{einstein05} the behavior
of the velocity components parallel and orthogonal to the relative
velocity between inertial systems, which is as close as one can get
without vectors to the vectorial version \eqref{eqEA}.

%%%%%%%%%%%%%%%%%%%%%%%%%%%%%%%%%%%%%%%%%%%%%%%%%%%%%%%%%%%%%%%%%%%%
%FIGURE 6-7: Einstein gyroline and cogyroline.      Deleted.
%%%%%%%%%%%%%%%%%%%%%%%%%%%%%%%%%%%%%%%%%%%%%%%%%%%%%%%%%%%%%%%%%%%%
%\input{fig174175f.eps} % The gyroline and cogyroline
%       in (+)E  Fig. 6-7 Fig.~\ref{fig174175fm}
%%% \begin{figure}
%%% \centerline{\includegraphics[width=10cm]{new_fig174175f.eps}}
%\caption{
%The gyroline (left) and cogyroline (right) in an
%Einstein gyrovector space of a disc
%are, respectively, a Euclidean straight line and an elliptical arc that
%intersects the boundary of the disc diametrically.
%The gyroline is recognized as the
%geodesic line in the Beltrami-Klein ball model of hyperbolic geometry
%\cite{ungardiff05}.
%\label{fig174175fm}}
%%% \end{figure}
%%%%%%%%%%%%%%%%%%%%%%%%%%%%%%%%%%%%%%%%%%%%%%%%%%%%%%%%%%%%%%%%%%%%
%\enlargethispage{16pt}

Seemingly structureless, Einstein velocity addition could not play
in Einstein's special theory of relativity a central role. Indeed,
Borel's attempt to ``repair'' the seemingly ``defective'' Einstein velocity
addition in the years following 1912 is described in
\cite[p.~117]{walter99b}.
Fortunately, however, there is no need to ``repair''
the Einstein velocity addition law since, like
M\"obius addition in the ball,
Einstein addition in the ball is
a gyrocommutative gyrogroup operation, which gives rise to the
Einstein ball gyrogroups $(\vs,\ope)$ and
gyrovector spaces $(\vs,\ope,\ode)$, Figs.~6\,--\,7 \cite{mybook01,sltwo01}.
Furthermore, Einstein's gyration turns out to be the Thomas precession of
relativity physics \cite{grouplike}, so that Thomas precession is a
kinematic effect rather than a dynamic effect as it is usually
portrayed \cite{ungarthomas06}.
A brief history of the discovery of Thomas precession
is presented in \cite[Sec.~1.1]{mybook01}.
 \enlargethispage{16pt}

The gamma factor is related to Einstein addition
by the {\it gamma identity}
\begin{equation} \label{eqgupv00}
\gupvbe = \gub\gvb\left(1+\frac{\ub\ccdot\vb}{s^2}\right)
\end{equation}
This gamma identity provided the historic link between
Einstein's special theory of relativity and the hyperbolic geometry of
Bolyai and Lobachevsky, as explained in \cite{ungareinstrigo07}.

Einstein scalar multiplication in the ball $\vs$ is identical
with M\"obius scalar multiplication, \eqref{eqmlt03}, in the ball $\vs$,
$r\ode\vb=r\odm\vb$ for all $r\inn\Rb$ and $\vb\inn\vs$.
Hence Einstein and M\"obius scalar multiplication are denoted here,
collectively, by $\od$.

The isomorphism between Einstein addition $\ope$ and
M\"obius addition $\opm$ in the ball $\vs$ is surprisingly simple
when expressed in gyrolanguage, the language of gyrovector spaces.
As we see from \cite[Table 6.1]{mybook02},
the gyrovector space isomorphism between
$(\vs,\ope,\od)$ and $(\vs,\opm,\od)$ is given by the equations
%%%%%%%%%%%%%%%%%%%%%%%%%%%%%%%%%%%%%%%%%%%%%%%%%%%%%%%%%%%%%%%%%%%%
 \begin{equation} \label{eq3ejksa}
 \begin{split}
\ub\ope\vb &= 2\od(\half\od\ub\opm\half\od\vb) \\[6pt]
\ub\opm\vb &= \half\od(2\od\ub\ope 2\od\vb)
 \end{split}
 \end{equation}
%%%%%%%%%%%%%%%%%%%%%%%%%%%%%%%%%%%%%%%%%%%%%%%%%%%%%%%%%%%%%%%%%%%%

Following \eqref{eqdfhn01}, Einstein cooperation, also called
Einstein coaddition, in the ball is
commutative, given by the equation
\begin{equation} \label{eqform28}
\ub \sqpe \vb = 2 \od \frac{\gub\ub+\gvb\vb}{\gub+\gvb}
\end{equation}
for all $\ub,\vb\inn\vs$.
Clearly, $\vb\sqme\vb=\zerb$.
Noting the {\it Einstein half},
\begin{equation} \label{idhjns}
\half\od\vb = \frac{\gvb}{1+\gvb}\vb
\end{equation}
and the {\it scalar associative law} of gyrovector spaces
\cite[p.~138]{mybook02},
it is clear from \eqref{eqform28}\,--\,\eqref{idhjns}
that $\vb\sqpe\zerb=\vb$, as expected.

Einstein noted in 1905 that
\begin{quotation}
``Das Gesetz vom Parallelogramm der Geschwindigkeiten gilt also nach
unserer Theorie nur in erster Ann\"aherung.''
\begin{flushright}
A.~Einstein \cite{einstein05}, 1905
\end{flushright}
\end{quotation}
[Thus the law of velocity parallelogram is valid according to our
theory only to a first approximation.]

We now see that with our gyrovector space approach to hyperbolic geometry,
Einstein's noncommutative addition $\ope$  gives rise to an {\it exact}
hyperbolic parallelogram addition $\sqpe$, Fig.~8,
which is commutative.
The cogyrogroup $(\vs,\sqp)$ is thus an important commutative loop
that regulates algebraically the hyperbolic parallelogram
\cite{gyroparallelogram06}.
%\enlargethispage{16pt}

An interesting contrast between Euclidean and hyperbolic geometry
is thus observed here.
In Euclidean geometry and in classical mechanics
vector addition coincides with
the parallelogram addition law.
In contrast, in hyperbolic geometry and in relativistic mechanics
gyrovector addition, given by Einstein addition,
$\ub\ope\vb$,
and the gyroparallelogram addition, $\ub\sqpe\vb$ in $\vs$,
are distinct.
We thus face the problem of whether the ultimate relativistic
velocity addition is given by the
%%%%%%%%%%%%%%%%%%%%%%%%%%%%%%%%%%%%%%%%%%%%%%%%%%%%%%%%%%%%%%%%%%%%%%%%%%%%%
%\begin{itemize}
%\item[$(i)$]
(i) non-commutative Einstein velocity addition law in \eqref{eqEA},
or by the
%\item[$(ii)$]
(ii) commutative Einstein gyroparallelogram addition law in
Fig.~8.
%\end{itemize}
%%%%%%%%%%%%%%%%%%%%%%%%%%%%%%%%%%%%%%%%%%%%%%%%%%%%%%%%%%%%%%%%%%%%%%%%%%%%%
Fortunately, a cosmic phenomenon that can provide the ultimate resolution
of the problem does exist.
It is the stellar aberration, illustrated
classically and relativistically
for particle aberration in Figs.~9 and 10.

%%%%%%%%%%%%%%%%%%%%%%%%%%%%%%%%%%%%%%%%%%%%%%%%%%%%%%%%%%%%%%%%%%%%
%FIGURE 8: Einstein gyroparallelogram
% Updated into Fig. 3
%%%%%%%%%%%%%%%%%%%%%%%%%%%%%%%%%%%%%%%%%%%%%%%%%%%%%%%%%%%%%%%%%%%%
 \input{fig190ips} % The Einstein gyroparallelogram. Fig.~\ref{fig190im}
%%%%%%%%%%%%%%%%%%%%%%%%%%%%%%%%%%%%%%%%%%%%%%%%%%%%%%%
% Input here the figure in "fig174e.eps"    Fig. 8       Updated: .eps -> .ps
%%%%%%%%%%%%%%%%%%%%%%%%%%%%%%%%%%%%%%%%%%%%%%%%%%%%%%%
%\begin{figure}
%\centerline{\includegraphics[width=10cm]{new_fig190i.eps}}
%%\caption{This is the caption of the figure.}
% \caption{
% The Einstein gyroparallelogram in an Einstein gyrovector space
%$(\vs,\op,\od)$, $\op=\ope$,
%and the Einstein gyroparallelogram addition law.
% \label{fig190im}}
%\end{figure}
%%%%%%%%%%%%%%%%%%%%%%%%%%%%%%%%%%%%%%%%%%%%%%%%%%%%%%%%%%%%%%%%%%%%

A cosmic experiment in our cosmic laboratory, the Universe, that can
validate the Einstein gyroparallelogram addition law,
Fig.~8, and its associated
gyrotriangle addition law of Einsteinian velocities shown in Fig.~10,
is the {\it stellar aberration} \cite{stewart64}.
Stellar aberration is particle aberration where the particle is a
photon emitted from a star.
Particle aberration, in turn, is the change in the apparent direction
of a moving particle caused by the relative motion between
two observers. The case when the two observers are E (at rest relative to
the Earth) and S (at rest relative to the Sun) is shown graphically in
Fig.~9 (classical interpretation)
and
Fig.~10 (relativistic interpretation).
Obviously, in order to detect stellar aberration there is no need to
place an observer at rest relative to the Sun since this effect
 varies during the year. It is this variation that can be
observed by observers at rest relative to the Earth.

The classical interpretation of particle aberration is obvious in terms
of the triangle law of Newtonian velocity addition
(which is the common vector addition in Euclidean geometry), as
demonstrated graphically in Fig.~9.
The relativistic interpretation of particle aberration is, however,
less obvious.

Relativistic particle aberration is illustrated in
%%%%%%%%%%%%%%%%%%%%%%%%%%%%%%%%%%%%%%%%%%%%%%%%%%%%%%%%%%%%%%%%%%%%
%FIGURES 9-10: Stellar Aberration.      Deleted.
%%%%%%%%%%%%%%%%%%%%%%%%%%%%%%%%%%%%%%%%%%%%%%%%%%%%%%%%%%%%%%%%%%%%
%\input{fig206sa2b2.eps} % Stellar Aberration.
%%% \begin{figure}
%%% \centerline{\includegraphics[width=10cm]{new_fig206sa2b2.eps}}
%\caption{
%\label{fig163164cm}}
%%% \end{figure}
%%%%%%%%%%%%%%%%%%%%%%%%%%%%%%%%%%%%%%%%%%%%%%%%%%%%%%%%%%%%%%%%%%%%
Fig.~10 in terms of analogies that it shares with its classical interpretation
in Fig.~9. These analogies are just analogies that gyrocommutative gyrogroups
share with commutative groups and gyrovector spaces share with
vector spaces. Remarkably, the resulting expressions that describe
the relativistic stellar aberration phenomenon, obtained by our
gyrovector space approach, agree with expressions that are obtained in the
literature by employing the relativistic Lorentz transformation group.
Our gyrovector space approach is thus capable of recovering known results
in astrophysics,
to which it gives new geometric interpretations that are analogous to
known, classical interpretations.

%SECTION 8
\section*{\centerline{8. Dark Matter of the Universe}}

\begin{quotation}
What is the universe made of? We do not know.
If standard gravitational theory is correct, then most of the matter in the
universe is in an unidentified form that does not emit enough light to have
been detected by current instrumentation. Astronomers and physicists are
collaborating on analyzing the characteristics of this dark matter and in
exploring possible physics or astronomical candidates for the unseen
material.
\begin{flushright}
S.~Weinberg and J.~Bahcall \cite[p.~v]{weinberg04}
\end{flushright}
\end{quotation}

Fortunately, our gyrovector space approach is capable of discovering
a novel result in astrophysics as well, proposing a viable mechanism
for the formation of the dark matter of the Universe.

We have seen in Sec.~8 that the cosmic effect of stellar aberration
supports our gyrovector gyrospace approach guided by analogies that it
shares with the common vector space approach.
Another cosmic effect that may support a relativistic physical
novel result obtained by our gyrovector space approach to
Einstein's special theory of relativity is related to the elusive
relativistic center of mass.
The difficulties in attempts to obtain a satisfactory
relativistic center of mass definition were discussed by
Born and Fuchs in 1940 \cite{bornfuchs40},
but they did not propose a satisfactory definition.
Paradoxically, ``In relativity, in contrast to Newtonian mechanics,
the centre of mass of a system is not uniquely determined'',
as Rindler stated with a supporting example \cite[p.~89]{rindler82}.
Indeed, in 1948 M.H.L. Pryce \cite{pryce48} reached the conclusion that
``there appears to be no wholly satisfactory definition of the
[relativistic] mass-centre.''
Subsequently, Pryce's conclusion was confirmed by many authors
who proposed various definitions for the relativistic center of mass;
see for instance
\cite{alba02,yaremko94,lehner95} and references therein,
where various approaches to the concept of the relativistic center of mass
are studied.
Consequently, Goldstein stated that
``a meaningful center-of-mass (sometimes called center-of-energy)
can be defined in special relativity only in terms of the
angular-momentum tensor, and only for
a particular frame of reference.'' \cite[p.~320]{goldstein}.

Fortunately, the spacetime geometric insight that our novel
grouplike loop approach offers enables the
elusive ``manifestly covariant'' relativistic center of mass of
a particle system with {\it proper time} to be identified.
It turns out to be analogous to the classical center of mass to the mass
of which a specified
fictitious mass must be added so as to render it
``manifestly covariant'' with respect to the motions of hyperbolic geometry.
Specifically, let $S = S(m_k,\vb_k,\Sigma_\zerb,N)$,
be an isolated system of
$N$ noninteracting material particles the $k$-th particle of which has
mass $m_k >0$ and velocity $\vb_k \inn\Rct$
relative to a rest frame $\Sigma_\zerb$,
$k=1,\dots,N$.
Then, classically, the system $S$ of $N$ particles can be viewed as a
fictitious single
particle located at the center of mass of $S$, with mass
$m_0 = \sum_{k=1}^{N} m_k$ that equals the total mass of the
constituent particles of $S$.
Relativistically, however, symmetries are determined by
gyrogroup, rather than group, symmetries.
As in the classical counterpart,
the system $S$ can be viewed in Einstein's special theory of relativity
as a fictitious single particle
located at the relativistic center of mass of $S$
(specified in \cite{mybook03}), with mass $m_0$ that we present
in \eqref{gksnw1} below.

In order to obey necessary relativistic symmetries,
the mass $m_0$ of the relativistic center of mass of $S$
must exceed, in general, the total mass of the
constituent particles of $S$ according to the equation
\begin{equation} \label{gksnw1}
m_0 \phantom{i} = \phantom{i} \sqrt{
\left( \sum_{k=1}^{N} m_k \right)^2 +
2\sum_{\substack{j,k=1\\j<k}}^N m_j  m_k
(\gamma_{\om\vb_j\op\vb_k}^{\phantom{O}} -1)
}
\phantom{i} \ge \phantom{i}
\sum_{k=1}^{N} m_k
\end{equation}
as explained in \cite{mybook03}.

The additional, fictitious mass $m_0 - \sum_{k=1}^{N} m_k$
in \eqref{gksnw1} of the system $S$
results from relative velocities, $\om\vb_j\op\vb_k$, $j,k=1,\ldots,N$,
between particles of the system $S$.
The fictitious mass of a rigid particle system, therefore, vanishes.
The fictitious mass of nonrigid galaxies does not vanish and, hence,
could account for the dark matter needed
to gravitationally ``glue'' each nonrigid galaxy together.

Indeed, the cosmic laboratory, our Universe, may support the existence of
the predicted fictitious mass in \eqref{gksnw1}
as the mass of the dark matter
in the Universe that astrophysicists are forced to postulate
but cannot detect \cite{weinberg04,milgrom02,cline03,nicolson07,trimble87}.
Hence, in order to uncover a viable mechanism that
accounts for the formation of dark matter
that manifests itself only through gravitational interaction,
there is no need to modify the laws of physics, as
Milgrom proposed in \cite{milgrom02}. Rather, one can find it in our
grouplike loop approach that improves our understanding of
Einstein's special theory of relativity and its underlying hyperbolic geometry
of Bolyai and Lobachevsky \cite{mybook03}.

%\newpage

%SECTION 9
\section*{\centerline{9. The Bloch Gyrovector of QIC}}

Bloch vector is well known in the theory of
quantum information and computation (QIC).
We will show that, in fact, Bloch vector is not a vector but, rather,
a gyrovector
\cite{bloch02,densityshort02,density02}.
It is easy to predict that in the present twenty-first century it is
quantum mechanics that will increasingly influence our lives.
Hence, it would be interesting
to see what gyrovector spaces have to offer in QIC.

A {\it qubit} is a two state quantum system completely described
by the qubit {\it density matrix} $\rho_\vb$,
\begin{equation} \label{eqwry03}
\rho_\vb
= \half
\begin{pmatrix} 1+ v_3 & v_1 - iv_2 \\ v_1 + iv_2 & 1- v_3 \end{pmatrix}
\end{equation}
parametrized by the vector $\vb=(v_1,v_2,v_3)\inn\BB^3$
in the open unit ball $\BB^3=\Rsthreeu$ of the
Euclidean 3-space $\Rt$. The vector $\vb$ in the ball is known in
QIC as the Bloch vector. However, we will see that it would be
more appropriate to call it a gyrovector rather than a vector.

The density matrix product of the four density matrices in the
following equation, which are parametrized by two distinct
Bloch vectors $\ub$ and $\vb$,
can be written as a single density matrix parametrized by
the Bloch vector $\wb$, multiplied by the trace of the matrix product,
\begin{equation} \label{eqweng04a}
\rho_\ub \rho_\vb \rho_\vb \rho_\ub =
tr[\rho_\ub \rho_\vb \rho_\vb \rho_\ub ]
\rho_\wb
\end{equation}
$\ub,\vb\inn\BB^3$.
Here $tr[m]$ is the trace of a square matrix $m$, and
\begin{equation} \label{eqweng05}
\wb=\ub\opm(2\od\vb\opm\ub)=2\od(\ub\opm\vb)
\end{equation}
Identity \eqref{eqweng05} is one of several identities available in
\cite{bloch02,densityshort02,density02} that demonstrate the
compatibility of density matrix manipulations and gyrovector space
manipulations.

Two Bloch vectors $\ub$ and $\vb$ generate the two density matrices
$\rho_\ub$ and $\rho_\vb$ that, in turn, generate the
{\it Bures fidelity}
${\cal F}(\rho_\ub,\rho_\vb)$
that we may also write as
${\cal F}(\ub,\vb)$.
The Bures fidelity
${\cal F}(\ub,\vb)$
is a most important distance measure between quantum states
$\rho_\ub$ and $\rho_\vb$ of the qubit in QIC,
given by the equations
\begin{equation} \label{eqwry04}
{\cal F}(\ub,\vb) =
\left[ {\rm tr} \sqrt{\sqrt{\rho_\ub} \rho_\vb \sqrt{\rho_\ub}} \right]^2
= \half \, \displaystyle\frac{
1 + \gamma_{\ub\ope\vb}^{\phantom{1}}
}{\gub\gvb}
\end{equation}
The first equation in \eqref{eqwry04} is well known
\cite{nielsen00,kwek00}, and the second equation in \eqref{eqwry04} is
a gyrovector space equation verified in \cite[Eq.~9.69]{mybook02}.
Identity \eqref{eqweng05} and
the second identity in \eqref{eqwry04} indicate that in density matrix
manipulations in QIC, Bloch vectors appear to behave like gyrovectors
in M\"obius gyrovector spaces $(\Rsthreeu,\opm,\od)$
and in Einstein gyrovector spaces $(\Rsthreeu,\ope,\od)$.

Indeed, since the Bures fidelity has particularly wide currency today in
QIC geometry, Nielsen and Chuang
had to admit for their chagrin \cite[p.~410]{nielsen00} that

\begin{quotation}
``Unfortunately, no similarly
[alluding to Euclidean geometric interpretation]
clear geometric interpretation is known for the fidelity between two states
of a qubit''.
%\begin{flushright}
%Nielsen and Chuang \cite[p.~410]{nielsen00}, 2000
%\end{flushright}
\end{quotation}

It is therefore interesting to realize that while Bures fidelity has no
Euclidean geometric interpretation, as Nielsen and Chuang admit,
it does have a hyperbolic geometric interpretation,
which is algebraically regulated by our grouplike loops
and their associated gyrovector spaces.

\noindent
\footnotesize{
         Department of Mathematics\\
         North Dakota State University\\
         Fargo, North Dakota 58105\\
         United States of America\\
e-mail: Abraham.Ungar@ndsu.edu}

\end{document}

%% file: fig174eps
% Fig. 193 gyroparallelepiped
\begin{figure}[t]  % try to put this figure on the top of the page
 \centering         % center the figure
\psfrag{pa}  {$A,\hspace{0.1cm}t=0$}
\psfrag{pb}  {$B,\hspace{0.1cm}t=1$}
\psfrag{formula01}[]{${\rm The~M\ddot obius~Gyroline}~L_{_{AB}}$}
\psfrag{formula02}[]{${\rm through~the~points}~A~{\rm and}~B$}
\psfrag{formula03}[]{$\boxed{ A\op(\om A\op B)\od t}$}
\psfrag{fig174m04}[]{$ -\infty < t < \infty $}
\includegraphics[width=9cm]{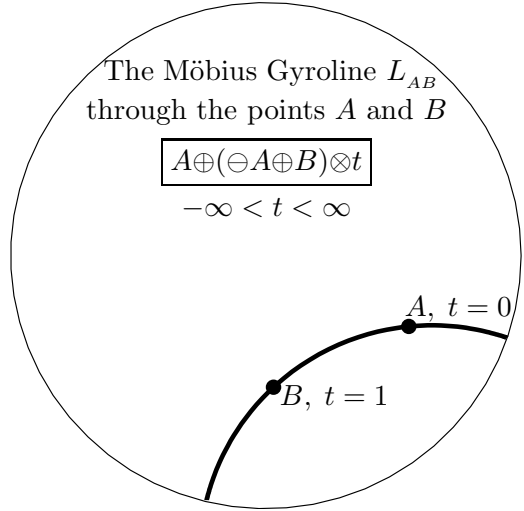}
 \caption[The M\"obius Gyroline]{
 In ``gyroformalism'',
 hyperbolic geometric expressions take the graceful forms of their
 Euclidean counterparts. This is convincingly illustrated by
 the unique gyroline in a M\"obius gyrovector plane $(\Rstwo,\op,\od)$
 through two given points $A$ and $B$ in the disc is shown.
 When the parameter $t\inn\Rb$ runs from $-\infty$ to $\infty$
 the point $p(t)=A\op(\om A\op B)\od t$ runs over the gyroline
 $L_{_{AB}}$. In particular, at ``time'' $t=0$ the point is at $p(0)=A$,
 and, owing to the left cancellation law of M\"obius addition,
 at ``time'' $t=1$ the point is at $p(1)=B$.
 The M\"obius gyroline equation is shown in the box. The analogies it shares
 with the Euclidean straight line equation in the vector space approach to
 Euclidean geometry are obvious.
 \label{fig174em}}
\end{figure}
%%%%%%%%%%%%%%%%%%%%%%%%%%%%%%%%%%%%%%%%%%%%%%%%%%%%%%%%%%%%%%%%%%%%%%%%%%%%%

%% file: fig148kps
\begin{figure}[t]  % try to put this figure on the top of the page
 \centering         % center the figure
%%%%%%%%%%%%%%%%%%%%%%%%%%%%%%%%%%%%%%%%%%%%%%%%%%%%%%%%%%%%%%%%%%%%
\psfrag{pointer01}[]{$\gyr[\pb(t_3),-\vbo]\ab$}
% The [] causes centering the LaTex box in the psfrag comand
% The [] can be deleted if not needed.
%%%%%%%%%%%%%%%%%%%%%%%%%%%%%%%%%%%%%%%%%%%%%%%%%%%%
\psfrag{paa}  {$\ab$}
\psfrag{pbb}  {$\bb$}
\psfrag{pcc}  {$\cb$}
\psfrag{pa}  {$A$}
\psfrag{pb}  {$B$}
\psfrag{pc}  {$C$}
\psfrag{f1}  {$\ab=\om C\op B$}
\psfrag{f2}  {$\bb=\om C\op A$}
\psfrag{f3}  {$\cb=\om B\op A$}
\psfrag{f4}  {$a=\|\ab\|,~b=\|\bb\|,~c=\|\cb\|$}
\psfrag{f5}  {$\frac{a^2}{s^2}
= \frac{\cos\alpha+\cos(\beta+\gamma)}{\cos\alpha+\cos(\beta-\gamma)}$}
\psfrag{f6}  {$\frac{b^2}{s^2}
= \frac{\cos\beta+\cos(\alpha+\gamma)}{\cos\alpha+\cos(\alpha-\gamma)}$}
\psfrag{f7}  {$\frac{c^2}{s^2}
= \frac{\cos\gamma+\cos(\alpha+\beta)}{\cos\gamma+\cos(\alpha-\beta)}$}
\psfrag{f8}[]{$\cos\gamma = \frac{\om C\op A}{\|\om C\op A}
                     \ccdot \frac{\om C\op B}{\|\om C\op B} $}
\psfrag{f9}[]{$\delta=\pi-(\alpha+\beta+\gamma)>0$}
%%%%%%%%%%%%%%%%%%%%%%%%%%%%%%%%%%%%%%%%%%%%%%%%%%%%
%%%%%%%%%%%%%%%%%%%%%%%%%%%%%%%%%%%%%%%%%%%%%%%%%%%%%%%%%%%%%%%%%%%%
\includegraphics[width=9cm]{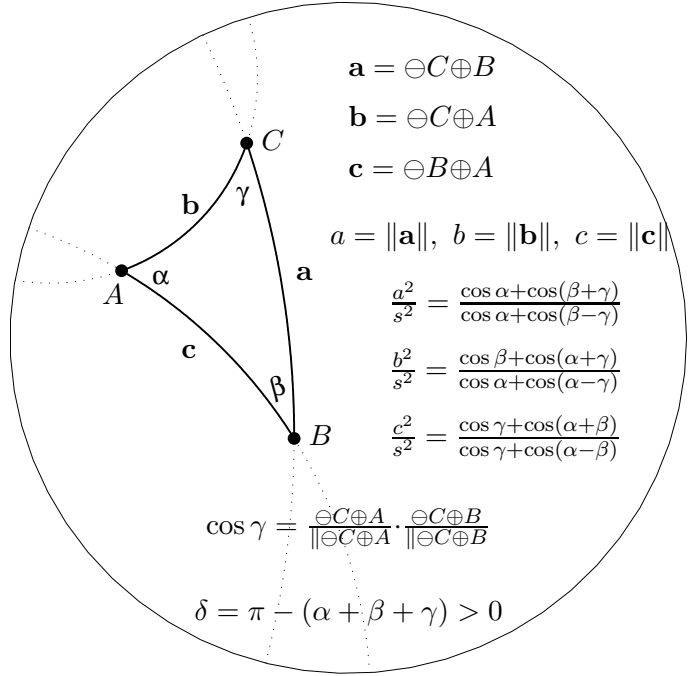}
\caption[]{
M\"obius gyrotriangle and its standard notation and identities
in a M\"obius gyrovector space $(\vs,\op,\od)$.
Remarkably, in the limit as $s\rightarrow\infty$
the equations in the figure reduce to their Euclidean counterparts.
Thus, for instance, in that limit we have
$\cos\alpha+\cos(\beta+\gamma)=0$
implying the Euclidean theorem
according to which the triangle angle sum is $\pi$,
$\alpha+\beta+\gamma=\pi$.
\label{fig148km}}
\end{figure}

%% file: fig190ips
 
%%%%%%%%%%%%%%%%%%%%%%%%%%%%%%%%%%%%%%%%%%%%%%%%%%%%%%%%%%%%%%%%%%%%%%
%%%%% The Einstein gyroparallelogram Addition Law   %%%%%%%%%%%%%%%%%%
%\begin{figure}[htbp]
\begin{figure}[t]  % try to put this figure on the top of the page
              % [h] tries to place the figure here
              % [b] tries to place the figure on the bottom of the page
              % [t] tries to place the figure on the top of the page
              % [P] tries to place the figure floatingly on the page
 \centering         % center the figure
\psfrag{----chets}[]{\lower-1.0ex \hbox {\footnotesize{$\blacktriangleright$}}}
\psfrag{A}[]{$A$}
\psfrag{B}[]{$B$}
\psfrag{C}[]{$\hspace{-.1cm}C$}
\psfrag{D}[]{$\hspace{0.2cm}D$}
\psfrag{u}[]{$\ub$}
\psfrag{v}[]{$\vb$}
\psfrag{w}[]{$\wb$}
\psfrag{formula00}{$\hspace{-0.6cm}D=(B\sqp C)\om A$}
\psfrag{formula01}{$\ub=\om A\op B$}
\psfrag{formula02}{$\vb=\om A\op C$}
\psfrag{formula03}{$\wb=\om A\op D$}
\psfrag{fig190i}[]{$\ub\sqp\vb=\wb$}
 \includegraphics[width=9cm]{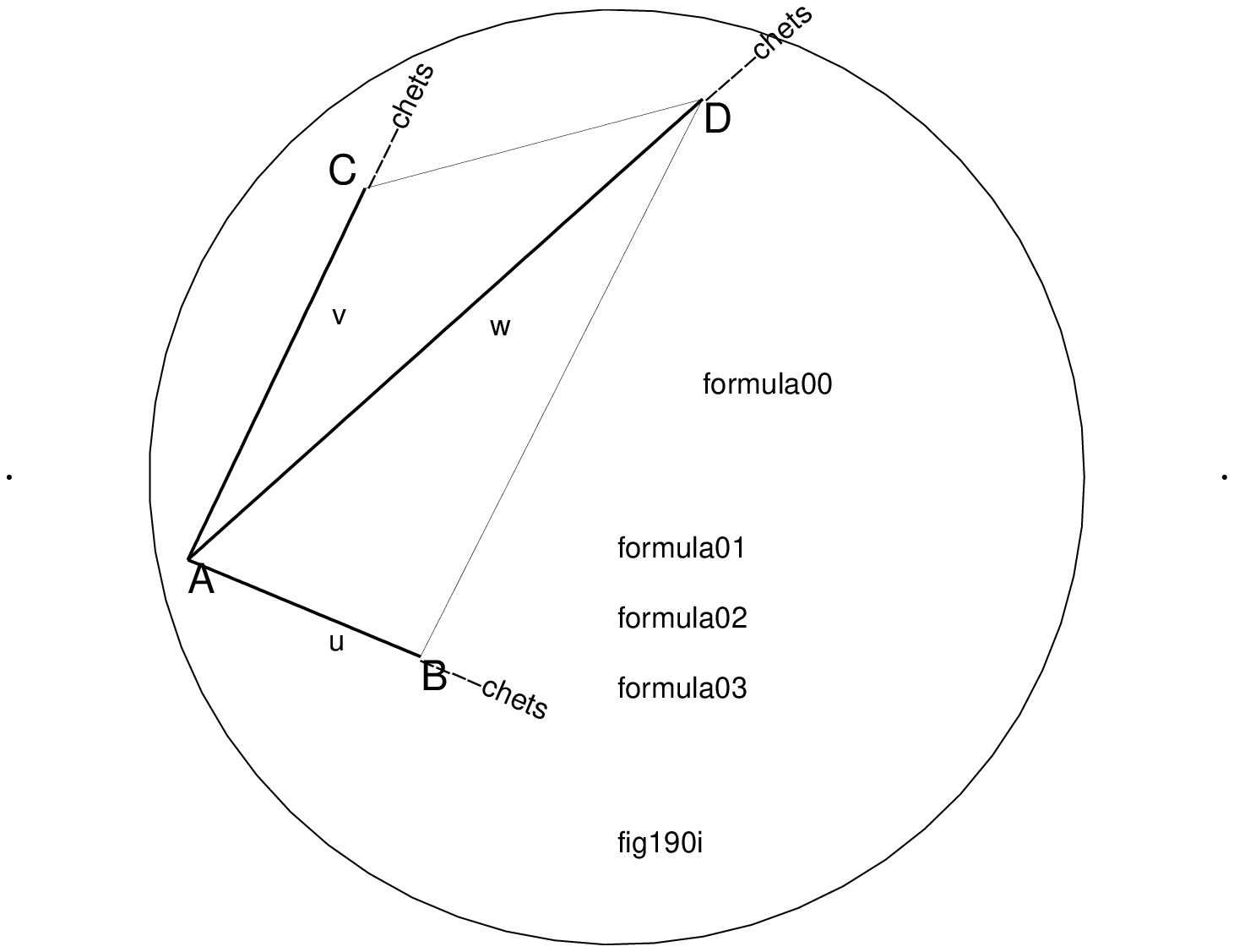}
\caption[The Einstein Gyroparallelogram Law]{
The Einstein gyroparallelogram addition law of
relativistically admissible velocities. 
Let $A,B,C\inn\Rst$ be any three nongyrocollinear points of an
Einstein gyrovector space $(\Rst,\op,\od)$, giving rise to the two
gyrovectors $\ub=\om A\op B$ and $\vb=\om A\op C$.
Furthermore, let $D$ be a point of the gyrovector space such that
$ABDC$ is a gyroparallelogram, that is, $D=(B\sqp C)\om A$.
Then, Einstein coaddition of
$\ub$ and $\vb$, $\ub\sqp\vb=\wb$, obeys the gyroparallelogram law,
$\wb=\om A\op D$, just as vector addition in $(\Rt,+)$ obeys the
parallelogram law. Einstein coaddition, $\sqp$, thus gives
rise to the gyroparallelogram addition law
of Einsteinian velocities, which is commutative and fully
analogous to the parallelogram addition law of Newtonian velocities.
\label{fig190im}}
\end{figure}
%%%%%%%%%%%%%%%%%%%%%%%%%%%%%%%%%%%%%%%%%%%%%%%%%%%%%%%%%%%%%%%%%%%%%%

%% file: ungar03_loops07.bbl
\small
\begin{thebibliography}{20}

\bibitem{ahlfors73}
Lars~V. Ahlfors.
\newblock {\em Conformal invariants: topics in geometric function theory}.
\newblock McGraw-Hill Book Co., New York, 1973.
\newblock McGraw-Hill Series in Higher Mathematics.

\bibitem{ahlfors81}
Lars~V. Ahlfors.
\newblock {\em M\"obius transformations in several dimensions}.
\newblock University of Minnesota School of Mathematics, Minneapolis, Minn.,
  1981.

\bibitem{alba02}
David Alba, Luca Lusanna, and Massimo Pauri.
\newblock Centers of mass and rotational kinematics for the relativistic
  {$N$}-body problem in the rest-frame instant form.
\newblock {\em J. Math. Phys.}, 43(4):1677--1727, 2002.

\bibitem{weinberg04}
John Bahcall, Tsvi Piran, and Steven Weinberg (eds.).
\newblock {\em Dark Matter in the Universe}.
\newblock sec.~ed. Kluwer Academic Publishers Group, Dordrecht, 2004.

\bibitem{barrett98}
J.~F. Barrett.
\newblock {\em Special relativity and hyperbolic geometry}.
\newblock Univ. Sunderland, Sunderland, UK, 1998.
\newblock Physical Interpretations of Relativity Theory. Proceedings, London,
  UK, 11--14 Sept. 1998.

\bibitem{gracielaungar01}
Graciela~S. Birman and Abraham~A.~Ungar.
\newblock The hyperbolic derivative in the poincar\'e ball model of hyperbolic
  geometry.
\newblock {\em J. Math. Anal. Appl.}, 254:321--333, 2001.

\bibitem{bornfuchs40}
M.~Born and K.~Fuchs.
\newblock The mass centre in relativity.
\newblock {\em Nature}, 145:587, 1940.

\bibitem{sltwo01}
Jing-Ling Chen and Abraham~A.~Ungar.
\newblock From the group ${\rm {s}{l}}(2,\bold {C})$ to gyrogroups and
  gyrovector spaces and hyperbolic geometry.
\newblock {\em Found. Phys.}, 31(11):1611--1639, 2001.

\bibitem{bloch02}
Jing-Ling Chen and Abraham~A.~Ungar.
\newblock The {B}loch gyrovector.
\newblock {\em Found. Phys.}, 32(4):531--565, 2002.

\bibitem{cline03}
David~B. Cline.
\newblock The search for dark matter.
\newblock {\em Sci.~Amer.}, March:50--59, 2003.

\bibitem{corry98}
Leo Corry.
\newblock The influence of {D}avid {H}ilbert and {H}ermann {M}inkowski on
  {E}instein's views over the interrelation between physics and mathematics.
\newblock {\em Endeavor}, 22(3):95--97, 1998.

\bibitem{einstein05}
Albert Einstein.
\newblock Zur {E}lektrodynamik {B}ewegter {K}\"orper [on the electrodynamics of
  moving bodies]
%({W}e use the {E}nglish translation in \cite{einsteinfive} or
%  in \cite{lorentz52}).
\newblock {\em Ann. Physik (Leipzig)}, 17:891--921, 1905.

%\bibitem{einsteinfive}
%Albert Einstein.
%\newblock {\em Einstein's Miraculous Years: Five Papers that Changed the Face
%  of Physics}.
%\newblock Princeton, Princeton, NJ, 1998.
%\newblock Edited and introduced by John Stachel. Includes bibliographical
%  references. Einstein's dissertation on the determination of molecular
%  dimensions -- Einstein on Brownian motion -- Einstein on the theory of
%  relativity -- Einstein's early work on the quantum hypothesis. A new English
%  translation of Einstein's 1905 paper on pp. 123--160.

\bibitem{feder03}
Tom{\'a}s Feder.
\newblock Strong near subgroups and left gyrogroups.
\newblock {\em J. Algebra}, 259(1):177--190, 2003.

\bibitem{fisher99}
Stephen~D. Fisher.
\newblock {\em Complex variables}.
\newblock Dover Publications Inc., Mineola, NY, 1999.
\newblock Corrected reprint of the second (1990) edition.

\bibitem{tuvalungar01}
Tuval Foguel and Abraham~A.~Ungar.
\newblock Involutory decomposition of groups into twisted subgroups and
  subgroups.
\newblock {\em J. Group Theory}, 3(1):27--46, 2000.

\bibitem{tuvalungar02}
Tuval Foguel and Abraham~A.~Ungar.
\newblock Gyrogroups and the decomposition of groups into twisted subgroups and
  subgroups.
\newblock {\em Pac. J. Math}, 197(1):1--11, 2001.

\bibitem{yaremko94}
R.~P. Ga{\u\i}da, V.~I. Tretyak, and Yu.~G. Yaremko.
\newblock Center-of-mass variables in relativistic {L}agrangian dynamics of a
  particle system.
\newblock {\em Teoret. Mat. Fiz.}, 101(3):402--416, 1994.

\bibitem{goldstein}
Herbert Goldstein.
\newblock {\em Classical mechanics}.
\newblock Addison-Wesley Publishing Co., Reading, Mass., second edition, 1980.
\newblock Addison-Wesley Series in Physics.

\bibitem{greenkrantz}
Robert~E. Greene and Steven~G. Krantz.
\newblock {\em Function theory of one complex variable}.
\newblock John Wiley \& Sons Inc., New York, 1997.
\newblock A Wiley-Interscience Publication.

\bibitem{harukirassias94}
Hiroshi Haruki and Themistocles~M. Rassias.
\newblock A new invariant characteristic property of {M}\"obius transformations
  from the standpoint of conformal mapping.
\newblock {\em J. Math. Anal. Appl.}, 181(2):320--327, 1994.

\bibitem{harukirassias96}
Hiroshi Haruki and Themistocles~M. Rassias.
\newblock A new characteristic of {M}\"obius transformations by use of
  {A}pollonius points of triangles.
\newblock {\em J. Math. Anal. Appl.}, 197(1):14--22, 1996.

\bibitem{harukirassias98}
Hiroshi Haruki and Themistocles~M. Rassias.
\newblock A new characteristic of {M}\"obius transformations by use of
  {A}pollonius quadrilaterals.
\newblock {\em Proc. Amer. Math. Soc.}, 126(10):2857--2861, 1998.

\bibitem{harukirassias00}
Hiroshi Haruki and Themistocles~M. Rassias.
\newblock A new characterization of {M}\"obius transformations by use of
  {A}pollonius hexagons.
\newblock {\em Proc. Amer. Math. Soc.}, 128(7):2105--2109, 2000.

\bibitem{hausner98}
Melvin Hausner.
\newblock {\em A vector space approach to geometry}.
\newblock Dover Publications Inc., Mineola, NY, 1998.
\newblock Reprint of the 1965 original.

\bibitem{nourou99}
A.~Nourou Issa.
\newblock Gyrogroups and homogeneous loops.
\newblock {\em Rep. Math. Phys.}, 44(3):345--358, 1999.

\bibitem{nouro01}
A.~Nourou Issa.
\newblock Left distributive quasigroups and gyrogroups.
\newblock {\em J. Math. Sci. Univ. Tokyo}, 8(1):1--16, 2001.

\bibitem{kiechle02}
Hubert Kiechle.
\newblock {\em Theory of ${K}$-loops}.
\newblock Springer-Verlag, Berlin, 2002.

\bibitem{kinyonjones00}
Michael~K. Kinyon and Oliver Jones.
\newblock Loops and semidirect products.
\newblock {\em Comm. Algebra}, 28(9):4137--4164, 2000.

\bibitem{kinyonungar00}
Michael~K. Kinyon and Abraham~A.~Ungar.
\newblock The gyro-structure of the complex unit disk.
\newblock {\em Math. Mag.}, 73(4):273--284, 2000.

\bibitem{kuznetsov03}
Eugene Kuznetsov.
\newblock Gyrogroups and left gyrogroups as transversals of a special kind.
\newblock {\em Algebra Discrete Math.}, (3):54--81, 2003.

\bibitem{lawson04}
Jimmie Lawson and Yongdo Lim.
\newblock Symmetric sets with midpoints and algebraically equivalent theories.
\newblock {\em Results Math.}, 46(1-2):37--56, 2004.

\bibitem{lehner95}
Luis~R. Lehner and Osvaldo~M. Moreschi.
\newblock On the definition of the center of mass for a system of relativistic
  particles.
\newblock {\em J. Math. Phys.}, 36(7):3377--3394, 1995.

%\bibitem{lorentz52}
%H.~A. Lorentz, A.~Einstein, H.~Minkowski, and H.~Weyl.
%\newblock {\em The principle of relativity}.
%\newblock Dover Publications Inc., New York, N. Y., undated.
%\newblock With notes by A. Sommerfeld, Translated by W. Perrett and G. B.
%  Jeffery, A collection of original memoirs on the special and general theory
%  of relativity.

\bibitem{marsden74}
Jerrold~E. Marsden.
\newblock {\em Elementary classical analysis}.
\newblock W. H. Freeman and Co., San Francisco, 1974.
\newblock With the assistance of Michael Buchner, Amy Erickson, Adam
  Hausknecht, Dennis Heifetz, Janet Macrae and William Wilson, and with
  contributions by Paul Chernoff, Istv\'an F\'ary and Robert Gulliver.

\bibitem{milgrom02}
Mordehai Milgrom.
\newblock Is 95\% of the universe really missing? an alternative to dark
  matter.
\newblock {\em Sci.~Amer.}, August:42--50, 2002.

\bibitem{wright02}
David Mumford, Caroline Series, and David Wright.
\newblock {\em Indra's pearls: The vision of Felix Klein}.
\newblock Cambridge University Press, New York, 2002.

\bibitem{needham97}
Tristan Needham.
\newblock {\em Visual complex analysis}.
\newblock The Clarendon Press Oxford University Press, New York, 1997.

\bibitem{nicolson07}
Iain Nicolson.
\newblock {\em Dark side of the universe: dark matter, dark energy, and the
  fate of the cosmos}.
\newblock John Hopkins University Press, Baltimore, MD, 2007.

\bibitem{nielsen00}
Michael~A. Nielsen and Isaac~L. Chuang.
\newblock {\em Quantum computation and quantum information}.
\newblock Cambridge University Press, Cambridge, 2000.

\bibitem{pflugfelder90}
Hala~O. Pflugfelder.
\newblock {\em Quasigroups and loops: introduction}, volume~7 of {\em Sigma
  Series in Pure Mathematics}.
\newblock Heldermann Verlag, Berlin, 1990.

\bibitem{pflugfelder00}
Hala~Orlik Pflugfelder.
\newblock Historical notes on loop theory.
\newblock {\em Comment. Math. Univ. Carolin.}, 41(2):359--370, 2000.
\newblock Loops'99 (Prague).

\bibitem{pryce48}
M.~H.~L. Pryce.
\newblock The mass-centre in the restricted theory of relativity and its
  connexion with the quantum theory of elementary particles.
\newblock {\em Proc. Roy. Soc. London. Ser. A.}, 195:62--81, 1948.

\bibitem{pyenson85}
Lewis Pyenson.
\newblock Relativity in late {W}ilhelmian {G}ermany: the appeal to a
  preestablished harmony between mathematics and physics.
\newblock {\em Arch. Hist. Exact Sci.}, 27(2):137--155, 1982.

\bibitem{ratcliffe94}
John~G. Ratcliffe.
\newblock {\em Foundations of hyperbolic manifolds}, volume 149 of {\em
  Graduate Texts in Mathematics}.
\newblock Springer-Verlag, New York, 1994.

\bibitem{rindler82}
Wolfgang Rindler.
\newblock {\em Introduction to special relativity}.
\newblock The Clarendon Press Oxford University Press, New York, 1982.

\bibitem{rozga00}
Krzysztof R{\'o}zga.
\newblock On central extensions of gyrocommutative gyrogroups.
\newblock {\em Pacific J. Math.}, 193(1):201--218, 2000.

\bibitem{sabinin95}
Lev~V.~Sabinin.
\newblock On {A}.~{U}ngar's gyrogroups.
\newblock {\em Uspekhi Mat.~Nauk}, 50(5(305)): 251--252, 1995.

\bibitem{sabinin98}
Lev~V.~Sabinin, Ludmila~L.~Sabinina, and Larissa~V.~Sbitneva.
\newblock On the notion of gyrogroup.
\newblock {\em Aequationes Math.}, 56(1-2):11--17, 1998.

\bibitem{stewart64}
Albert~B. Stewart.
\newblock The discovery of stellar aberration.
\newblock {\em Sci.~Amer.}, March: 100--108, 1964.

\bibitem{trimble87}
V.~Trimble.
\newblock Existence and nature of dark matter in the universe.
\newblock {\em Ann. Rev. Astron. Astrophys.}, 25:425--472, 1987.

\bibitem{parametrization}
Abraham~A.~Ungar.
\newblock Thomas rotation and the parametrization of the {L}orentz
  transformation group.
\newblock {\em Found. Phys. Lett.}, 1(1):57--89, 1988.

\bibitem{noncomm}
Abraham~A.~Ungar.
\newblock The relativistic noncommutative nonassociative group of velocities
  and the {T}homas rotation.
\newblock {\em Resultate Math.}, 16(1-2):168--179, 1989.
\newblock The term ``K-loop'' is coined here.

\bibitem{grouplike}
Abraham~A.~Ungar.
\newblock Thomas precession and its associated grouplike structure.
\newblock {\em Amer. J. Phys.}, 59(9):824--834, 1991.

\bibitem{mybook01}
Abraham~A.~Ungar.
\newblock {\em Beyond the {E}instein addition law and its gyroscopic {T}homas
  precession: The theory of gyrogroups and gyrovector spaces}, volume 117 of
  {\em Fundamental Theories of Physics}.
\newblock Kluwer Academic Publishers Group, Dordrecht, 2001.

\bibitem{densityshort02}
Abraham~A.~Ungar.
\newblock The density matrix for mixed state qubits and hyperbolic geometry.
\newblock {\em Quantum Inf. Comput.}, 2(6):513--514, 2002.

\bibitem{density02}
Abraham~A.~Ungar.
\newblock The hyperbolic geometric structure of the density matrix for mixed
  state qubits.
\newblock {\em Found. Phys.}, 32(11):1671--1699, 2002.

\bibitem{mybook02}
Abraham~A.~Ungar.
\newblock {\em Analytic hyperbolic geometry: Mathematical foundations and
  applications}.
\newblock World Scientific Publishing Co. Pte. Ltd., Hackensack, NJ, 2005.

\bibitem{ungardiff05}
Abraham~A.~Ungar.
\newblock Gyrovector spaces and their differential geometry.
\newblock {\em Nonlinear Funct. Anal. Appl.}, 10(5):791--834, 2005.

\bibitem{ungarthomas06}
Abraham~A.~Ungar.
\newblock Thomas precession: a kinematic effect of the algebra of {E}instein's
  velocity addition law. {C}omments on: ``{D}eriving relativistic momentum and
  energy. {II}. {T}hree-dimensional case'' [{E}uropean {J}. {P}hys. {\bf 26}
  (2005), no. 5, 851--856; mr2227176] by {S}. {S}onego and {M}. {P}in.
\newblock {\em European J. Phys.}, 27(3):L17--L20, 2006.

\bibitem{gyroparallelogram06}
Abraham~A.~Ungar.
\newblock The relativistic hyperbolic parallelogram law.
\newblock In {\em Geometry, integrability and quantization}, pages 249--264.
  Softex, Sofia, 2006.

\bibitem{ungareinstrigo07}
Abraham~A. Ungar.
\newblock Einstein's velocity addition law and its hyperbolic geometry.
\newblock {\em Comput. Math. Appl.}, 53(8):1228--1250, 2007.

\bibitem{mbtogyp07}
Abraham~A.~Ungar.
\newblock From {M}\"obius to gyrogroups.
\newblock {\em Amer. Math. Monthly}, 2007.
\newblock in print.

\bibitem{mybook03}
Abraham~A.~Ungar.
\newblock {\em Analytic hyperbolic geometry and Einstein's special theory
  of relativity}.
\newblock World Scientific Publishing Co. Pte. Ltd., Hackensack, NJ, 2008.
\newblock in print.

\bibitem{vermeer05}
J.~Vermeer.
\newblock A geometric interpretation of {U}ngar's addition and of gyration in
  the hyperbolic plane.
\newblock {\em Topology Appl.}, 152(3):226--242, 2005.

\bibitem{walter99a}
Scott Walter.
\newblock Minkowski, mathematicians, and the mathematical theory of relativity.
\newblock In {\em The expanding worlds of general relativity (Berlin, 1995)},
  pages 45--86. Birkh\"auser Boston, Boston, MA, 1999.

\bibitem{walter99b}
Scott Walter.
\newblock The non-{E}uclidean style of {M}inkowskian relativity.
\newblock In {\em The symbolic universe (J. J. Gray (ed.), Milton Keynes,
  England)}, pages 91--127. Oxford Univ. Press, New York, 1999.

\bibitem{walterrev2002}
Scott Walter.
\newblock Book {R}eview: {\it {B}eyond the {E}instein {A}ddition {L}aw and its
  {G}yroscopic {T}homas {P}recession: {T}he {T}heory of {G}yrogroups and
  {G}yrovector {S}paces}, by {A}braham {A}. {U}ngar.
\newblock {\em Found. Phys.}, 32(2):327--330, 2002.

\bibitem{kwek00}
Xiang-Bin Wang, L.~C. Kwek, and C.~H. Oh.
\newblock Bures fidelity for diagonalizable quadratic {H}amiltonians in
  multi-mode systems.
\newblock {\em J. Phys. A}, 33(27):4925--4934, 2000.

\end{thebibliography}
